%% file: main.tex
\documentclass[letter,9pt]{article}
\usepackage[scale=.83]{geometry}
\usepackage[affil-it]{authblk}
\usepackage{amssymb,amsmath,amsthm,amsbsy}
\usepackage{physics}
\usepackage{graphicx}
\usepackage{epstopdf}
\usepackage{epsfig}
\usepackage{multicol}
\usepackage{graphicx}
\usepackage{caption}
\usepackage{multirow}
\usepackage{algorithm,algorithmic}

\usepackage{setspace}
\usepackage{color}

\usepackage{lineno}
\usepackage{svg}
\usepackage{pdfpages}
\usepackage{subfig}
\usepackage{comment}
\usepackage{cleveref}
\usepackage{booktabs}
\usepackage{nomencl}
\makenomenclature
\usepackage{ifthen}
\usepackage{microtype} 

\usepackage{titlesec} 
\usepackage[toc]{appendix}
\titleformat{\section}[block]{\large\scshape\centering}{\thesection.}{1em}{} 
\titleformat{\subsection}[block]{\large}{\thesubsection.}{1em}{} 

\title{\textbf{A Reduced Order Model for Joint Assemblies by Hyper-Reduction and Model-Driven Sampling}}

\author{Ahmed Amr Morsy\footnote{Corresponding author.\\ \normalfont{E-mail address: morsya@ethz.ch}}\ \textsuperscript{,\textdagger}}
\author{Mariella Kast\footnote{Both authors contributed equally to this work.}}
\author{Paolo Tiso}
\affil{
	Institute for Mechanical Systems, ETH Zürich, Leonhardstrasse 21, 8092 Zurich, Switzerland}

\date{}

\input{commands}

\begin{document}
	
\graphicspath{{figures/}}

\maketitle 
\thispagestyle{empty}

\noindent\textbf{Abstract}: The dynamic behavior of jointed assemblies exhibiting friction nonlinearities features amplitude-dependent dissipation and stiffness. To develop numerical simulations for predictive and design purposes, macro-scale High Fidelity Models (HFMs) of the contact interfaces are required. However, the high computational cost of such HFMs impedes the feasibility of the simulations. To this end, we propose a model-driven method for constructing hyper-reduced order models of such assemblies. Focusing on steady-state analysis, we use the Multi-Harmonic Balance Method (MHBM) to formulate the equations of motion in frequency domain. The reduction basis is constructed through solving a set of vibration problems corresponding to fictitious interface conditions. Subsequently, a Galerkin projection reduces the order of the model. Nonetheless, the necessary fine discretization of the interfaces represents a bottleneck for achieving high speedups. For this reason, we implement an adapted Energy Conserving Weighing and Sampling (ECSW) technique for Hyper Reduction (HR), thereby allowing significant speedups for meshes of arbitrary fineness. This feature is particularly advantageous since analysts typically encounter a trade-off between accuracy and computational cost when deciding on the mesh size, whose estimation is particularly challenging for problems of this type. To assess the accuracy of our method without resorting to the HF solution, we propose an error indicator with thresholds that have proven reliable in our analyses. Finally, the accuracy and efficiency of the method are demonstrated by two case studies.

\section{Introduction}
Friction occurs on contact interfaces of joints, which are commonly found in mechanical, civil and aerospace engineering applications. The dynamic response of jointed assemblies can be heavily influenced by the nonlinearities of frictional contact resulting in amplitude-dependent dissipation and stiffness \cite{Brake2018b}. As demands increase on optimization for performance and cost, including minimizing experimental testing for calibration of models during design phases, predictive numerical simulations of joints become increasingly important. Any attempt to model friction inevitably depends on the length and time scales considered, as well as the purpose of the analysis. For instance, regarding the time scale, Fantetti et al. \cite{Fantetti2019} have recently presented insightful experimental evidence demonstrating the time evolution of both the coefficient of friction and the tangential stiffness. A significant increase in both parameter values during an initial number of fretting cycles, followed by a steady-state stabilization of the parameters has been observed. Yuan et al. \cite{Yuan2021a} have built upon the experimental results reported in \cite{Fantetti2019}, using the statistical distribution of the  stabilized values from the different tests, to perform a sensitivity analysis with respect to the contact parameters for steady-state experimental and numerical analysis of an underplatform damper system. The variability of the friction coefficient was shown to significantly affect the amplitude of the responses which featured considerable slipping along the interfaces, while the impact of the variability of the tangential stiffness was predominate for responses that featured low amount of slipping. The time scale we consider for our analysis corresponds to the steady-state behavior of the system, which is assumed to emerge after an initial number of fretting cycles.

As for the length scale, our work focuses on the accurate prediction of friction-induced dissipation, which demands in turn the capacity to simulate the microslip behavior in joints \cite{Gaul1997}. Microslip refers to the isolated, partial slipping of the interfaces relative to each other, in contrast to gross slip, which is the complete slipping of the interfaces. Different modelling strategies exist to reproduce/predict the microslip behavior. A recent review of damping models for friction has been presented in \cite{Mathis2020}, where connections between different damping models were highlighted. For predictive analyses, Macro-scale High-Fidelity Models (HFMs) which involve dense meshes of nonlinear elements, are widely used. In this context, modelling of frictional contact differs depending on assumptions as to whether, the tangential motion on the interface is one-dimensional or two-dimensional, coupling is assumed for tangential planar motion, and if contact pressure temporally evolves across the interface. Firrone and Zucca compared features of the responses obtained by employing different formulations in  \cite{Maria2011}. Brake et al. \cite{Brake2017} showed experimentally that a correct simulation of joint behavior must take into account time varying contact pressure across the interface. An interfacial mesh whose elements can reproduce this effect, equipped with an elastic or rigid Coulomb law, makes the depiction of global microslip behavior possible \cite{Krack2017}. Yet, despite the accuracy of such HFMs, a major drawback is their high computational cost, which stands in the way of efficient, predictive analyses.

Projection-based Reduced Order Models (ROMs) reduce the size of dynamical systems by projecting them onto suitable low-dimensional subspaces, thus providing accurate and efficient solutions. An efficient family of methods used extensively in substructuring is Component Mode Synthesis (CMS) methods \cite{Hurty1965}, which include the widely common Hurty/Craig-Bampton (HCB) method \cite{Craig1968}. In fact, many ROM techniques aimed for frictional contact problems implement a CMS first step, by which the degrees of freedom (DOFs) are reduced into fixed-interface modes and physically-retained boundary DOFs, which include interfacial DOFs. A Galerkin projection of the system onto the subspace defined by the transformation matrix is then carried out. HCB provides an example of a model-driven ROM, as opposed to a data-driven ROM, since the construction of the reduction basis is based solely on the model and no data from HF simulations is required. Due to the high number of retained interfacial DOFs in the case of HFM of joints, further reductions are necessary. Therefore, attempts have been made to supplement CMS methods using the concept of modal derivatives\cite{Idelsohn1985,Slaats1995}. The concept, which has been implemented for problems featuring geometric nonlinearity \cite{Wu2016}, has been adapted for friction problems by Pichler et al. \cite{Pichler2017a}. In their work, they formulate Trial Vector Derivatives (TVDs) for time domain analysis and use them for test cases where the normal contact, in particular, is accorded predominant importance. A comparison of the efficiency of different contact and friction models has also been presented in \cite{Pichler2017}. Hughes et al. \cite{Hughes2021} proposed an interface reduction method where they use System-level Characteristic Constraint (SCC) modes and Properly Orthogonal Interface Modal Derivatives (POIMDs). Their results show that global quantities can be reliably assessed using SCC modes with impressive speedups reaching to the order of thousands, while a high accuracy on local quantities such as slipping areas is possible with the addition of POIMDs, with overall speedups ranging between 10-20.

In parallel, ROM techniques for frequency domain analysis focusing on steady-state analysis have been developing. Krack presented the concept of extended periodic motion \cite{Krack2015}, which is essential to his extension of nonlinear modes to friction-damped systems \cite{Krack2014}. Using this method, the dynamics of the system are assumed to take place on a manifold that is constituted of a union of periodic motions coming from the computed nonlinear mode. The external forcing and linear damping are subsequently superimposed to the ROM. The method proved accurate and efficient for the test cases in \cite{Krack2014}, where reasonable assumptions regarding linear damping and isolated resonance are met. Mitra et al. \cite{Mitra2016} proposed the Amplitude Microslip Projection (AMP) method, where the construction of the reduction basis is based on eigensolutions at linearized configurations which are obtained by artificially probing the contact interface. Gastaldi et al. \cite{Gastaldi2018} presented the Jacobian Projection (JP) technique, which extends AMP to the multi-harmonic context, accounting for static-dynamic coupling. Along this line, we present an Augmented JP method, whose augmented vectors have proven necessary for accurate detection of local quantities such as stresses on the interface, as well as global amplitude-dependent dissipation and frequency particularly in a case where multiple contact interfaces are present. Yet, in spite of the reduced size of a ROM, a typical problem in case of HF modelling of joints is the large number of nonlinear elements, which in turn demands expensive computations in full-space to compute the nonlinear forces. Therefore, attention is directed to Hyper Reduction (HR) techniques that target reducing the computational cost of the nonlinear forces. 

Balaji et al. \cite{Balaji2021} have presented a ROM strategy through HR of the interfaces. In their work, HR refers to a re-presentation of the contact interface such that the computation of nonlinear forces is based solely on the reduced DOFs. They present two strategies: a novel whole-joint formulation that is representative of the HF model, where the interface is divided into patches based on level sets of a field objective such as the contact pressure after a static step, and another strategy by which coarser re-meshing of the interface with contours of a chosen field objective are used to place new nodes on the interface. Their results captured modal characteristics using a system of significantly reduced size. Still in the spirit of reducing nonlinear computations, Yuan et al. \cite{Yuan2020} present an adapted component mode synthesis method where the idea is to reduce the interface DOFs by adapting the reduction basis based on the nodes that are expected to actually exhibit nonlinear behavior, i.e. not sticking. Their novel approach is conceptually in line with continuation methods from the standpoint of optimizing the choice of initial conditions as the selection of (nonlinear) reduced interface DOFs is updated at each new frequency point.  Speedups ranging from 50-100 have been reported considering the cost per iteration for a test case with relatively restrictive assumptions regarding contact conditions. A further improvement was presented in \cite{Yuan2021} that excludes numerical noise, as well as nodes that are dominated by sticking behavior, from influencing the choice of reduced interface DOFs. To contrast our work with existing methods, we regard that relative to \cite{Balaji2021}, we explore employing a HR technique that retains the original interfacial mesh, and in relation to \cite{Yuan2020,Yuan2021}, we test considering only one reduced set of elements for the nonlinear computations at all frequency points. 

(Discrete) Empirical Interpolation Method (D)EIM \cite{Chaturantabut2010} and Energy Conserving Sampling and Weighting (ECSW) \cite{Remacle2012} are popular techniques to approximate nonlinear terms which require long computation times. In particular, the ECSW method approximates the nonlinear forces via energy-based quantities, where the respective weights are computed from HF snapshots of the forces \cite{Remacle2012}. If the ROM is data-driven, the HF force snapshots required for the generation of training vectors will present no additional costs. On the other hand, for a model-driven ROM, the idea is to avoid costly HF solutions. Therefore, for instance, in the context of geometrically nonlinear problems, it was proposed to perform a lifting of the linear modal subspace onto a quadratic manifold, thereby eliminating the need for HF snapshots \cite{Jain2018}. To cover a wider class of geometric nonlinearities, the training generation proposed later in \cite{Rutzmoser2017} comes from solutions to nonlinear static systems where the applied forcing belongs to Krylov subspace and is stochastically produced. In our work, we present an adaptation of ECSW aimed at friction nonlinearities, where the training generation presents no additional costs with respect to the ROM construction. Moreover, it is well-known that the determination of the mesh size for frictional contact problems is a daunting task and convergence studies with respect to interfacial mesh size are rarely carried out \cite{Krack2017}. To this end, we will show that such studies become completely within reach using our method.
 
\begin{figure}
\centering
\includegraphics[scale=0.30]{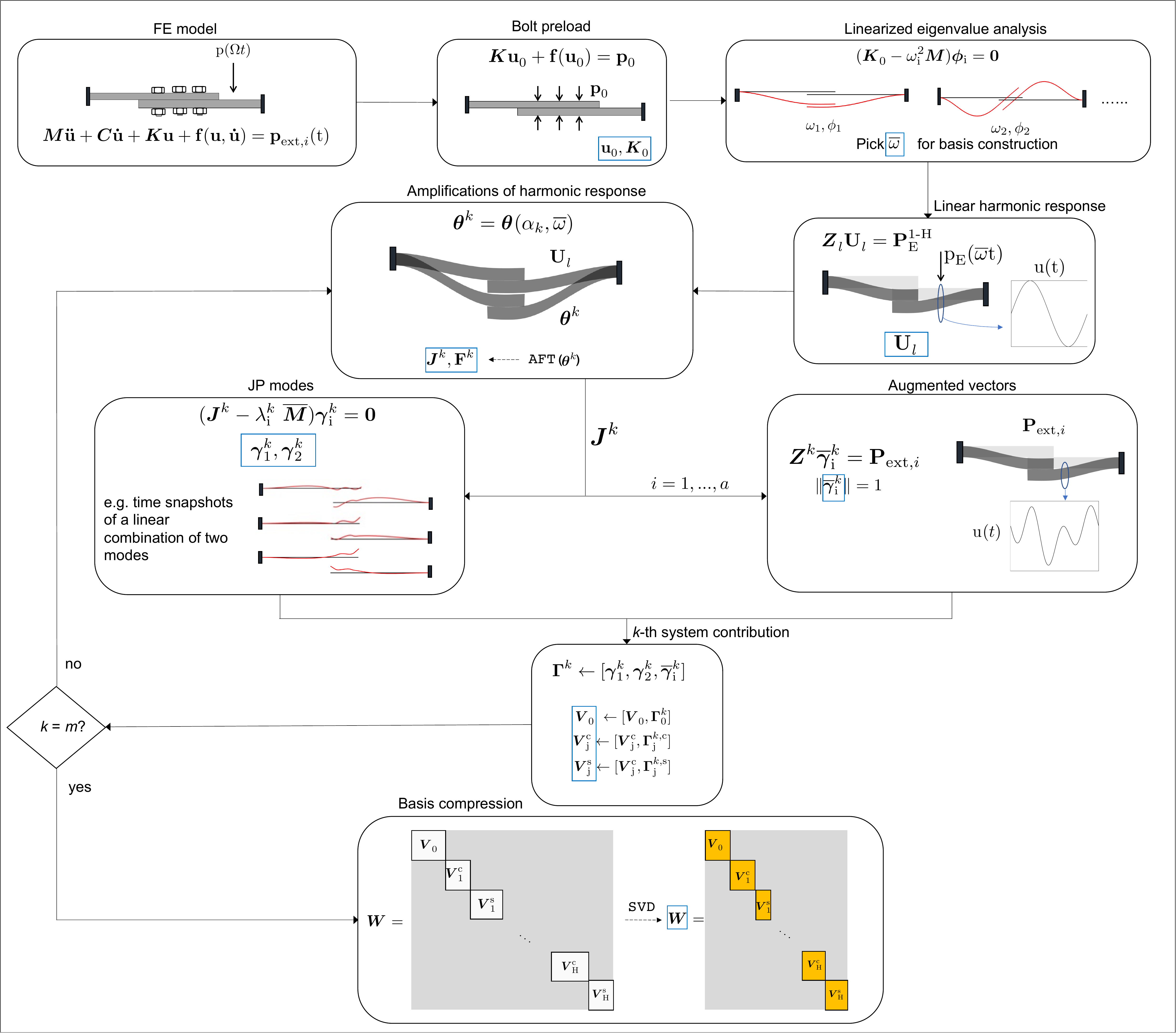}
\caption{Flowchart of the method 1/2 - Basis Construction at $\Omega = \overline{\omega}$.}
\label{fig:flowchart_basis}
\end{figure}

The paper is organized as follows: in Section \ref{subsection:formulation}, we present the mathematical formulation of the problem, along with the adopted contact law for frictional contact, and the computation of contact displacement is illustrated. An overview of frequency-domain analysis using MHBM is presented in Section \ref{sec:MHBM}. Next, we express two typical preliminary computations in Section \ref{Sec:preliminary_comp}. Subsequently, we describe the Augmented JP ROM strategy in \ref{subsection:reduced_basis}, then review ECSW HR in Section \ref{subsection:hyper_reduction_original} and adapt it to frictional contact in  Section \ref{subsec:hyper_reduction_training}. Figures \ref{fig:flowchart_basis} and \ref{fig:flowchart_HR} depict a complete flowchart of the method. In Section \ref{subsec:error_metric}, we present an error indicator that depends only on the results of HR-ROM. The accuracy and efficiency of the method is tested on two case studies in sections \ref{sec:case_study_beam} and \ref{sec:case_study_frame}. Finally, we present the conclusions in Section \ref{sec:conclusions}.

\section{Joint modelling framework for vibrations}
\subsection{Problem Formulation}
\label{subsection:formulation}
The momentum balance of a structural continuum consisting of an assembly of jointed structural components, discretized using Finite Element Method (FEM), can be expressed as

\begin{equation}
     \mat{M}\vec{\ddot{u}}\text{(t)} + \mat{C}\vec{\dot{u}}\text{(t)} + \mat{K}\vec{u}\text{(t)} + \vec{f}  (\vec{u}\text{(t)},\vec{\dot{u}}\text{(t)}) = \vec{p}_{\text{ext,}i}\text{(t)}
     \label{eq:FEDynEquil}
\end{equation}
with
\begin{equation}
 \vec{p}_{\text{ext,}i} \text{(t)}= \vec{p}_{\text{0}} + \alpha_{i} \vt{p}{E}{}(\Omega \text{t})
\label{eq:external_forces}
\end{equation}
where $\vec{u}\text{(t)} \in \mathbb{R}^{n} $ is the vector of nodal displacements, with $n$ being the number of Degrees of Freedom (DOFs). The time dependence of the kinematic variables in \eqref{eq:FEDynEquil} will be dropped from now on for clarity. We assume linear elasticity and small displacements. Matrices $\mat{K}, \mat{C},\mat{M}  \in \mathbb{R}^{n\times n}$  are the "free" stiffness, damping, and mass matrix of the mechanical system, which means that no coupling effects are considered in the assembly of the matrices due to the different components.  The vector of external forces $\vec{p}_{\text{ext,i}} \in \mathbb{R}^{n}$ is split into a vector of static forces $\vec{p}_{\text{0}}$, including possibly bolt preclamping forces, and a periodic forcing of unit value in $\vec{p}_{\text{E}}\ (\Omega \text{t})$, with an excitation frequency $\Omega$, amplified with a set of $a$ amplitudes of excitation $\alpha_{i}$ where $i=1,...,a$.

\begin{figure} [h!]
\centering
\includegraphics[scale=0.6]{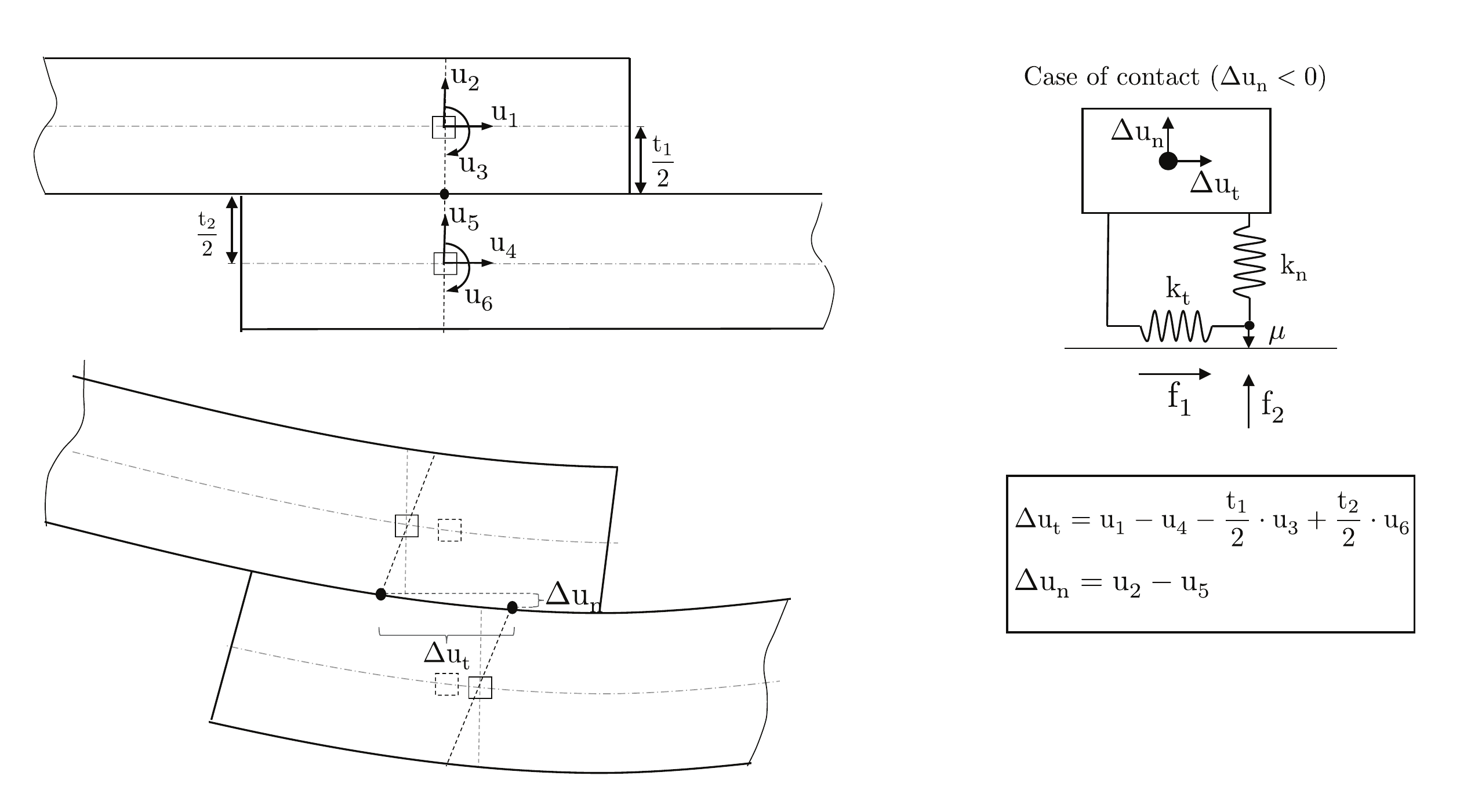}
\caption{Tangential and normal forces in case of contact}
\label{fig:contact_lawl}
\end{figure}

The vector $\vec{f} (\vec{u},\vec{\dot{u}})  \in \mathbb{R}^{n}$ represents the modelled frictional contact forces. Since we are studying joints in the microslip regime, we carry over the assumption of small displacements to contact kinematics, therefore contact is assumed to occur between fixed nodal pairs on the interface during vibrations. Accordingly, we adopt the node-to-node Jenkins element formulation with a unilateral spring in the normal direction \cite{Petrov2003}. The normal contact force $\st{f}{n}{}$ is given by

\begin{equation}
    \st{f}{n}{} = \begin{cases}
      \st{k}{n}{} \Delta\st{u}{n}{}  & \text{for penetration}, \\
      0 & \text{for separation},
    \end{cases}
    \label{eq:contact_law_1}
\end{equation}
where $\Delta \st{u}{n}{}$ is the relative normal displacement of the contact nodal pair, $\st{k}{n}{}$ is normal contact stiffness. As for the tangential force,

\begin{equation}
    \st{f}{t}{} = \begin{cases}
      \st{k}{t}{}(\Delta\st{u}{t}{} -\Delta\st{u}{t}{st}) + \st{f}{t}{st}  & \text{for sticking,} \\
      \mu \st{f}{n}{} \zeta  & \text{for slipping}, \\
      0  & \text{for separation},
    \end{cases}
    \label{eq:contact_law_2}
\end{equation}
where $\st{f}{t}{}$ is the tangential force, $\st{k}{t}{}$ is the tangent contact stiffness, $\Delta \st{u}{t}{st}$ and $\st{f}{t}{st}$ refer to the relative displacement and the friction force between two interface nodes at the start of the current stick phase. Regarding the slipping case, the Coulomb friction coefficient is denoted by $\mu$, and $\zeta=\text{sgn}(\st{f}{t}{slip})$ represents the sign of  $\st{f}{t}{slip}$, which is the tangential force at the end of the previous stick phase \cite{Petrov2003}. The transition between sticking and slipping is thus defined by a slip limit force, which depends on the normal contact force $\abs{\mu  \st{f}{n}{}}$. Note that despite the non-smooth evolution of contact forces in time due to the possibility of separation and slipping behaviors, the contact forces remain both piece-wise linear and $C^{0}$ continuous in time. Figure \ref{fig:contact_lawl} illustrates the computation of relative displacements, and represents the contact law in case of contact.

Physically, this model allows for three contact behaviors on a local level: sticking, slipping and lift-off. For small relative tangential displacements between the node pairs, the friction force is elastic, whereas slippage between the nodes, such as depicted by the original Coulomb type friction, is assumed to occur once a threshold of displacement is reached depending on the normal contact force. The modelled friction force satisfies the Masing hypothesis leading to a closed hysteresis curve for periodic motions\cite{Masing1923, Jayakumar1987}. The full contact force vector $\vec{f} (\vec{u})$ is constructed by taking into account the behavior of all Jenkins elements across the contact interfaces, which leads  to a nonlinear, non-smooth, spatial variation of forces. 

\subsection{Overview of Multi-Harmonic Balance Method}
\label{sec:MHBM}
The dynamic response of mechanical systems can be analyzed in time or frequency domains. For time domain analysis, time integration schemes are employed to transform the system of ODEs in \eqref{eq:FEDynEquil} into a set of nonlinear algebraic equations to be solved at each time step. If one is interested in investigating a periodic response under steady-state conditions due to one excitation frequency, a large number of time steps is required until the transient response decays, especially when structural damping is low. Alternatively, one can efficiently study the effect of multiple excitation amplitudes across a wide range of frequencies by resorting to a frequency domain approach. In particular, we implement MHBM, which we review in this section.

To compute a periodic solution of \eqref{eq:FEDynEquil}, the steady-state displacement solution $\vec{u}\text{(t)}$ is approximated using Fourier base functions as

\begin{equation}
\label{eq:QAnsatz}
    \vec{u}_{\text{h}}(\text{t}) = \vec{U}_0 + \sum_{\text{j}=1}^{\text{H}} \left({\vec{U}_\text{j}^{\text{c}} \cos(\text{j} \Omega \text{t}) + \vec{U}_\text{j}^{\text{s}} \sin(\text{j}\Omega \text{t})} \right),
\end{equation}
where the vector $\vec{U}_0  \in \mathbb{R}^{n}$ contains the static coefficients, H is the chosen number of harmonics, and the vectors $\vt{U}{c}{j}, \vt{U}{s}{j}  \in \mathbb{R}^{n}$ contain the Fourier coefficients of the cosine and sine components of the j-th harmonic, respectively. These components are collected in a column vector $\vec{U}$ as $\vec{U}= [\vt{U}{0}{}\ \vt{U}{1}{c}\ \vt{U}{1}{s}\ ...\ \vt{U}{H}{c}\  \vt{U}{H}{s}]^{\text{T}}$. Similarly, the nonlinear forces are approximated as

\begin{equation}
    \vt{f}{h}{}(\text{t}) = \vt{F}{0}{} + \sum_{\text{j}=1}^\text{H} \left({\vt{F}{j}{c} \cos({\text{j}\Omega \text{t}}) + \vt{F}{j}{s} \sin {(\text{j}\Omega \text{t}}) } \right),
    \label{FourierNonlinForce}
\end{equation}
where

\begin{equation}
\label{eq:FouriercoeffsF}
\begin{split}
\vt{F}{0}{} &= \dfrac{1}{\text{T}} \int_{0}^{\text{T}} \vec{f}\vec{(u)} \dd  \text{t} ,\\
\vt{F}{j}{c} &= \dfrac{2}{\text{T}} \int_{0}^{\text{T}} \vec{f}\vec{(u)}\cos({\text{j} \Omega \text{t}}) \dd  \text{t}, \\
\vt{F}{j}{s} &= \dfrac{2}{\text{T}} \int_{0}^{\text{T}}  \vec{f}\vec{(u)}\sin({\text{j} \Omega \text{t}} )\dd  \text{t}.
\end{split}
\end{equation}

However, the closed-form evaluation of the integrals in \eqref{eq:FouriercoeffsF} is complicated due to the coupling of the different harmonics, and the diverse phase transitions that an element potentially undergoes during one time period. Therefore, we employ the Alternating-Frequency Time (AFT) scheme \cite{Cameron1989,Cardona1994}, which can be summarized as follows: given the displacement vector $\vec{U}$ in the frequency-domain, one reconstructs $\vt{u}{h}{}(\text{t})$ across a sampled time period using \eqref{eq:QAnsatz}. Subsequently, the nonlinear forces are evaluated at the sampled time instants using the adopted contact law. One can then use a FFT to obtain the Fourier coefficients of the contact force vector in \eqref{eq:FouriercoeffsF}. 

MHBM refers to a weighted residual formulation of the equations of motion that are now represented using the Ansatz, where the residual forces  are imposed to be orthogonal to the weight functions (Fourier base functions) in a weak sense (see \cite{Krack2019}, for instance, for more details). Consequently, one obtains
\begin{equation}
\label{eq:NLQsys}
\mat{Z}\vec{U} + \vec{F} (\vec{U}) - \vec{P}_{\text{ext,}i} = \vec{0},
\end{equation}
where $\vec{F}, \vec{P}_{\text{ext,}i} \in \mathbb{R}^{n(2H+1)}$ are vectors collecting all Fourier coefficients of the nonlinear forces and the external forces as
\begin{equation}
\vec{F}= [\vt{F}{0}{}\ \vt{F}{1}{c}\ \vt{F}{1}{s}\ ...\ \vt{F}{H}{c}\  \vt{F}{H}{s}]^{\text{T}}, \\
\label{eq:vec_freq}
\end{equation}
\begin{equation}
\vec{P}_{\text{ext,}i}= \vt{P}{0}{}\ +\alpha_{i} \vt{P}{E}{}
\label{eq:p_ext_freq}
\end{equation}
where
\begin{equation}
\begin{split}
\vt{P}{0}{}= [\vt{P}{0}{}\ \vt{0}{}{}\ \vt{0}{}{}\ ...\ \vt{0}{}{}\  \vt{0}{}{}]^{\text{T}},\\
\vt{P}{E}{}= [\vt{0}{}{}\ \vt{P}{E,1}{c}\ \vt{P}{E,1}{s}\ \vt{0}{}{}\ ...\ \vt{0}{}{}\  \vt{0}{}{}]^{\text{T}},\\
\end{split}
\label{eq:vec_freq_comp}
\end{equation}
and $\mat{Z}$ is a blockdiagonal matrix, with block matrices $\mt{Z}{j}{} \in \mathbb{R}^{n \times n}$:
\begin{equation}
\label{eq:Zmatrix}
    \mat{Z} = \begin{bmatrix}
    \mt{Z}{0}{}   &   & & &\\
    &     \mt{Z}{1}{} & & & \\
    & & ... & & &\\
    & & & \mt{Z}{j}{} && \\
     & & & & ..& \\
    & & &  &&\mt{Z}{h}{}
    \end{bmatrix},
\end{equation}
where 
$\mt{Z}{0}{}= \mat{K}$, and

\begin{equation}
\label{eq:Z_diag_term}
\mt{Z}{j}{} = \mt{Z}{MC,j}{} (\Omega)+ \mt{Z}{K}{}=\begin{bmatrix}
-(\text{j} \Omega)^2 \mat{M} & \text{j} \Omega \mat{C} \\
-\text{j} \Omega \mat{C} & -(\text{j} \Omega)^2 \mat{M}
\end{bmatrix}
+\begin{bmatrix}
\mat{K} & \mat{0} \\
\mat{0} & \mat{K}
\end{bmatrix}
\end{equation}

We refer to the solution of \eqref{eq:NLQsys} as the High Fidelity (HF) solution. Note that due to the nonlinearity of \eqref{eq:NLQsys}, the solution is sought using an iterative Newton-Raphson (N-R) scheme:

\begin{equation}
\begin{split}
\label{eq:NR}
\vec{R}(\vec{U}^{(r)}) = \mat{Z}\vec{U}^{(r)} &+ \vec{F} (\vec{U}^{(r)}) - \vec{P}_{\text{ext,}i}\\\
\vec{R}(\vec{U}^{(r+1)}) \approx \vec{R}(\vec{U}^{(r)}) &+ \left.\dfrac{\pd \vec{R}(\vec{U}) }{\pd\vec{U}}\right|_{\vec{U}^{(r)}} \left({\vec{U}^{(r+1)}-\vec{U}^{(r)}}\right) = \vec{0}, \\
\end{split}
\end{equation}
where superscript $^{(r)}$ denotes the r-th iteration, and
\begin{equation}
\dfrac{\pd \vec{R}(\vec{U}) }{\pd\vec{U}}  = \mat{Z} + \dfrac{\pd\vec{F} (\vec{U})} {\pd\vec{U}},
\nonumber
\end{equation}
where the evaluation of the matrix $\dfrac{\pd\vec{F} (\vec{U})} {\pd\vec{U}} $ is carried out using the AFT procedure, with the derivatives analytically computed in the time domain, then transformed to the frequency domain using the FFT \cite{Cardona1994,Suess2015}. For the first iteration of \eqref{eq:NR}, we use as an initial guess the converged solution of the previous frequency point in the frequency sweep. As for the very first iteration, we use a linear dynamic solution as an initial guess, as we discuss in the next section.

\section{Hyper Reduced ROM for Joints}
\subsection{Preliminary computations}
\label{Sec:preliminary_comp}

\begin{itemize}
    \item \emph{Modal analysis of preclamped configuration}\\
    In order to determine the approximate frequency of the mode shape of interest of the jointed structure, we first solve the nonlinear static step:

\begin{equation}
\label{eq:NLstaticeq}
    \mat{K}
    \vt{u}{0}{}  +\vec{f} (\vt{u}{0}{}) = \vt{p}{0}{},
\end{equation}

where $\vec{u_\text{0}}  \in \mathbb{R}^{n}$ is the solution to the displacements under the effect of preclamping forces $\vt{p}{0}{}$. Having computed the static configuration $\vec{u_\text{0}}$, we define a stiffness matrix $\mt{K}{\emph{l}}{}$ that is linearized around $\vt{u}{0}{}$, which now takes into account the contact states at the interface, as

\begin{equation}
\label{eq:J_lin}
     \mt{K}{\emph{l}}{} = \mat{K} + \dfrac{\pd\vec{f} (\vt{u}{}{})} {\pd\vec{u}} \bigg\rvert_{\st{u}{0}{}}
\end{equation}

Next, we perform a modal analysis of the linearized, preclamped structure by the solving the eigenvalue problem of size $n$:
\begin{equation}
\label{eq:mode_freq}
(\mt{K}{\emph{l}}{} - \omega^{2}_{\text{i}} \mat{M}) \vec{\phi_{\text{i}}} = \vec{0}.    
\end{equation}
From the set of eigenfrequencies $\omega_\text{i}$, we choose that corresponding to the vibration mode of interest, and denote it by $\overline{\omega}$.
    
\item \emph{Linearized dynamic solution}\\
We compute the dynamic part of the response of the linearized structure using

\begin{equation}
\label{eq:LinDyn_freq}
    \mt{Z}{\emph{l}}{}
    \vt{U}{\emph{l}}{} = \vt{P}{E}{1-H},\\
\end{equation}
where 1-H refers to retaining only the dynamic components of the vector $\vt{P}{E}{}$ such that $ \vt{P}{E}{1-H} \in \mathbb{R}^{n(2H)}$. Vector $\vt{U}{\emph{l}}{} \in \mathbb{R}^{n(2H)}$ is the displacement due to the unitary force vector $ \vt{P}{E}{1-H} \in \mathbb{R}^{n(2H)}$, while the matrix $\mt{Z}{\emph{l}}{} \in \mathbb{R}^{n(2H) \times n(2H)}$ now consists of diagonal blocks, where the j-th block written as
\begin{equation}
\mt{Z}{\emph{l},j}{}= \mt{Z}{MC,j}{} (\Omega)+ \begin{bmatrix}
\mt{K}{\emph{l}}{} & \mat{0}  \\
\mat{0} & \mt{K}{\emph{l}}{}
\end{bmatrix}.
\label{eq:Z_diag_term}
\end{equation}

Note that regardless of the number of harmonics $\text{H}$ in the MHBM Ansatz, the dynamic response $\vt{U}{\emph{l}}{}$ only features one harmonic, due to the block diagonal structure of $\mt{Z}{\emph{l}}{}$ \eqref{eq:Z_diag_term}. Expressing \eqref{eq:LinDyn_freq} and \eqref{eq:Z_diag_term} is therefore chosen for convenience and clarity later on. Appending the static component $\vt{u}{0}{}$ \eqref{eq:NLstaticeq}, and accounting for an amplitude of excitation $\alpha_{i}$, the total linear response  becomes

\begin{equation}
\vt{\theta}{}{}(\alpha_{i},\Omega)=[\vt{u}{0}{T}\ \alpha_{i} \vt{U}{\emph{l}}{}(\Omega) ^{\text{T}}].
\label{eq:theta}
\end{equation}
where $\vt{\theta}{}{}=\vt{\theta}{}{}(\alpha_i,\Omega_1)$ is used to compute the initial guess for the first frequency $\Omega_{1}$ for a frequency sweep corresponding to the i-th amplitude. Moreover, \eqref{eq:theta} is used with $\vt{\theta}{}{}=\vt{\theta}{}{}(\alpha_{i},\overline{\omega})$ in the construction of the reduction basis as discussed in the next section.

\end{itemize}

\subsection{Augmented Jacobian Projection Method}
\label{subsection:reduced_basis}
Although jointed assemblies exhibit non-smooth dynamics, the adopted contact law remains piecewise linear. This feature is leveraged in the JP method \cite{Gastaldi2018}, since the contribution of each nonlinear element to the jacobian stiffness matrix becomes straightforward to determine, for given interface condition. Conceptually, the JP method is based on solutions of free vibration problems of a set of simulated linear systems, whose eigenvectors, in aggregate, are assumed to well approximate the response of the nonlinear system in steady-state condition. 

One starts by assuming a trial vector that reflects an expected baseline configuration of the interface. Next, this trial vector is amplified to simulate different boundary conditions on the interface, leading to systems with diverse contact states. Gastaldi et al. \cite{Gastaldi2018} suggest using amplifications of the stuck mode, which refers to a state where the interfaces are fully in contact, and are in sticking condition\footnote{In case the constructed basis results in a ROM containing high errors, they suggest using a computed displacement solution as a new trial vector to construct a new basis.}. For our purposes, we construct $m$ linear solutions of \eqref{eq:theta}, corresponding to $m$ amplitudes $\alpha_{k=1,...m}$, computed at resonance, and denote each by $\vec{\theta}^{k}$, such that
\begin{equation}
\vec{\theta}^{k}=\vt{\theta}{}{}(\alpha_{k},\overline{\omega}).
\label{eq:amplif}
\end{equation}
where $k=1,...m$. Even though our default strategy is to set $m=a$ , where $a$ is the number of excitation amplitudes in \eqref{eq:external_forces}, an enrichment of the basis with $m>a$ may be required in some cases as will be shown in Section \ref{sec:case_study_frame}. Next, and in line with \cite{Gastaldi2018}, once we choose the amplified displacement vector, it is then used as an input to the AFT algorithm to compute a nonlinear force vector $\vec{F}^{k}$, and the nonlinear stiffness contribution $\dfrac{\pd\vec{F}^{k} (\vec{U})} {\pd\vec{U}}$, for each contact state associated to $\vec{\theta}^{k}$. Note that while the input features dynamics consisting only of one harmonic component, the outputs will feature higher harmonic contributions due to the anticipated slipping and separation stimulated by the amplified trial vectors $\vec{\theta}^{k}$. Next, an eigenvalue problem describing the free vibrations of each intermediary, linear, autonomous system is now posed as

\begin{align}
\label{eq:eigen_lin}
(\mat{J}^{k} - \lambda_{\text{i}}^{k}\  \overline{\mat{M}}) \vec{\gamma}_{\text{i}}^{k} = \vec{0}, \\
\nonumber \norm{\vec{\gamma}_{\text{i}}^{k}}= 1,
\end{align}
where

 \begin{equation}
     \mat{J}^{k}= \mat{K}{}{}+\dfrac{\pd\vec{F} (\vec{U}^{k})} {\pd\vec{U}},
     \label{eq:jacobian}
 \end{equation}
and

\begin{equation}
\label{eq:Mmatrix}
\overline{\mat{M}} = 
\begin{bmatrix}
     
    \mat{0}   &   & & & &\\
    &     \mt{M}{}{} &  & & & & &&&\\
    &  &   \mt{M}{}{}&   & & &&&&\\
    
    & & & \ddots &  & &&&&\\
    &  &  &&  \st{j}{}{2}\mt{M}{}{}&   & & &&\\
    &  & &  && \st{j}{}{2}\mt{M}{}{}&   &  &&\\
    & & & &&& \ddots &  & &\\
    
    & & & && && H^{2}\mt{M}{}{}  & &\\
    & & & & & &&& H^{2}\mt{M}{}{}&  
\end{bmatrix}
=
\begin{bmatrix}
\mat{0}   & \mat{0}\\
\mat{0} & \overline{\mat{M}}_\text{0,H}
\end{bmatrix},
\end{equation}
with $\mat{J}^{k}, \overline{\mat{M}} \in \mathbb{R}^{n(2H+1) \times n(2H+1)}$. To avoid the ill-conditioning of \eqref{eq:eigen_lin} that is associated to the first zero block diagonal term of $\overline{\mat{M}}$, a static condensation step is performed (superscript $k$ is dropped for clarity):

\begin{equation}
\label{eq:static_cond1}
\begin{bmatrix}
\mat{J}_\text{0,0} & \mat{J}_\text{0,H} \\
\mat{J}_\text{H,0} & \mat{J}_\text{H,H} 
\end{bmatrix}
\begin{bmatrix}
\vt{\gamma}{i,0}{} \\
\vt{\gamma}{i,H}{}  
\end{bmatrix}
-
\lambda_i
\begin{bmatrix}
\mat{0} & \mat{0} \\
\mat{0} & \overline{\mat{M}}_\text{H,H} 
\end{bmatrix}
\begin{bmatrix}
\vt{\gamma}{i,0}{}  \\
\vt{\gamma}{i,H}{}  
\end{bmatrix}
=\vec{0}
\end{equation}

\begin{equation}
\label{eq:staticretrieval}
\mat{J}_\text{0,0}\  \vt{\gamma}{i,0}{}  + \mat{J}_\text{0,H}\  \vt{\gamma}{i,H}{}  = \vec{0} \Rightarrow \vt{\gamma}{i,0}{} = - (\mat{J}_\text{0,0}) ^{-1}\ \mat{J}_\text{0,H}\ \vt{\gamma}{i,H}{} \\
\end{equation}
\begin{equation}
\left(\mat{J}_\text{H,H} - \left( \mat{J}_\text{H,0}\  \mat{J}^{-1}_\text{0,0}\  \mat{J}_\text{0,H}\right) \right)\ \vt{\gamma}{i,H}{} -\ \lambda_i\ \overline{\mat{M}}_\text{H,H}\ \vt{\gamma}{i,H}{} = \vec{0}. 
\label{eq:static_cond2}
\end{equation}
Since the approximate frequency $\overline{\omega}$ of the mode shape of interest is already available from  \eqref{eq:mode_freq}, we select from \eqref{eq:static_cond2} the pair of eigenvectors $\vt{\gamma}{1,H}{}$, $\vt{\gamma}{2,H}{}$ with frequencies $\omega_{1}$, $\omega_{2}$ closest to $\overline{\omega}$. Note that we select two eigenvectors because we are interested in a response that is dominated by one harmonic (e.g fundamental harmonic), and the number of modes is 2$n$, owing to the cosine/sine representation. In case of complex eigenvectors, as could be already anticipated by the absence of symmetry in $\mat{J}^{k}$ in case slipping and/or separation occur, we only need the real and imaginary parts of one complex mode to construct the corresponding two-dimensional invariant subspace in $\mathbb{R}^{n(2H)}$ for the $k$-th linear system. Subsequently, we group the two modes as 
\begin{equation}
    \mat{\Gamma}_\text{H}^{k} = [\vec{\gamma}_\text{1,H}^{k}\ ,  \vec{\gamma}_\text{2,H}^{k}\ ],
    \label{eq:modes_dynamic}
\end{equation}
where $\mat{\Gamma}_\text{H}^{k} \in \mathbb{R}^{n(2H) \times 2} $. As for the static part of the eigenvectors, it should be highlighted that the JP method \cite{Gastaldi2018} takes into account the coupling between static and dynamic components in the off-diagonal blocks $\mt{J}{0,H}{}$, $\mt{J}{H,0}{}$. Therefore, we compute the static components of the eigenvectors using \eqref{eq:staticretrieval}, and group them as

\begin{equation}
    \mat{\Gamma}_{0}^{k} = [\vec{\gamma}_{1,0}^{k}\ ,  \vec{\gamma}_{2,0}^{k}\ ],
    \label{eq:modes_static}
\end{equation}
where $\mat{\Gamma}_{0}^{k} \in \mathbb{R}^{n \times 2} $. Assembling the components in one matrix $\mat{\Gamma}^{k}$, we write

\begin{equation}
\mat{\Gamma}^{k} =
  \begin{bmatrix}
  \mat{\Gamma}_{0}^{k}\\ \mat{\Gamma}_\text{H}^{k}
  \end{bmatrix},
  \label{eq:Gamma_matrix}
\end{equation}
where $\mat{\Gamma}^{k} \in \mathbb{R}^{n(2H+1) \times 2} $. In comparison to the original method \cite{Gastaldi2018}, we augment the JP basis $\mat{\Gamma}^{k}$ by computing the displacement solutions of the forced systems

\begin{align}
\label{eq:augment}
  \mat{Z}^{k} \vec{\overline{\gamma}}_\text{i}^{k}=  \vec{P}_{\text{ext,}i},
\end{align}
where $i=1,...,a$, and $\vec{P}_{\text{ext,}i} \in \mathbb{R}^{n(2H+1)}$ (see \eqref{eq:p_ext_freq}). Matrix $\mat{Z}^{k} \in \mathbb{R}^{n(2H+1) \times n(2H+1)}$  involves cross-harmonic coupling in the stiffness term, unlike  \eqref{eq:Z_diag_term}, and consists of:
\begin{equation}
    \mat{Z}^{k}= 
    \begin{bmatrix}
\mat{0}   & \mat{0}\\
\mat{0} & \mt{Z}{MC}{}
\end{bmatrix}
+ \mat{J}^{k}.
\end{equation}
The vectors are normalized as:
\begin{equation}
    \norm{\vec{\overline{\gamma}}_\text{i}^{k}}=1.
\end{equation}
Note that the LU decomposition of the dense matrix $\mat{Z}^{k}$ in the linear system \eqref{eq:augment}
 should be carried out only once for each k-system. We then augment the matrix $\mat{\Gamma}^{k}$ with $\vec{\overline{\gamma}}_\text{i}^{k}$. In fact, these augmented vectors have proven necessary for an enhanced accuracy on the forces such that that an error metric on the residual forces, such as what we present in Section \ref{subsec:error_metric}, could be used on the test cases. In addition, it has proven essential as well for an increased accuracy on the displacements for the more complicated test cases in Section \ref{sec:case_study_frame}.

Next, and similar to \cite{Gastaldi2018}, we apply this procedure for $k=1,...,m$ amplifications to construct $m$ matrices $\vec{\Gamma}^{k}$ that are then assembled to form $\mt{\Gamma}{}{}$ (see Algorithm \ref{algoritm:basis}). Then, $\mat{\Gamma}$ is partitioned according to its harmonic components. Finally, a Singular Value Decomposition (SVD) procedure is performed on each basis to avoid ill-conditioning, thereby the reduction basis for $\mt{V}{j}{c/s}$ for each harmonic component is constructed. 

As a result, each harmonic component can be represented as

\begin{equation}
    \vt{U}{j}{c/s} \approx \mt{V}{j}{c/s} \vt{Q}{j}{c/s},
\end{equation}
or, for the whole system, as

\begin{equation}
    \vec{U} \approx \mat{W} \vec{Q},
\end{equation}
where
$$\mat{W} = \begin{bmatrix} \mt{V}{0}{} &&&&&\\
&\mt{V}{1}{c} &&&&\\
&&\mt{V}{1}{s} &&& \\
&&&\dots&&\\
&&&& \mt{V}{H}{c}&\\
&&&&&\mt{V}{H}{s}
\end{bmatrix} .$$
Matrix $\mat{W} \in \mathbb{R}^{n(2H+1) \times p}$ has $p$ total number of columns in $\mat{W}$. Note that the bases $\mt{V}{0}{} \in \mathbb{R}^{n \times m_{0}}$, $\mt{V}{1}{c} \in \mathbb{R}^{n \times m_{1c}}$, etc., have in general different number of columns. \footnote{We note that an alternative strategy would be to construct $\mat{W}$ as a full matrix, in which there is no longer an association between a set of reduced DOFs and a distinct harmonic component. In our experience, this has resulted in a low accuracy of the associated ROM. A possible explanation is that the SVD procedure reduces then the contributions of higher harmonics inaccurately, since they typically have small magnitudes.} The reduced representation of the displacements in time domain can now be expressed as
\begin{equation}
\begin{split}
     \vt{u}{h}{} (\text{t}) & =  \vt{U}{0}{} + \sum_{\text{j}=1}^{\text{H}} \left({\vt{U}{j}{c} \cos(\text{j} \Omega \text{t}) + \vt{U}{j}{s} \sin {(\text{j} \Omega \text{t}}) } \right) \approx \vt{V}{0}{} \vt{Q}{0}{} + \sum_{\text{j}=1}^{\text{H}} \left(\vt{V}{j}{c}{\vt{Q}{j}{c} \cos({\text{j} \Omega \text{t}}) + \vt{V}{j}{s} \vt{Q}{j}{s} \sin {(\text{j} \Omega \text{t}}) } \right). \label{eq:reduced_disp}
\end{split}
\end{equation}
Next, the subspace of $\mat{W}{}{}$ is used to construct the ROM via the Galerkin projection:

\begin{equation}
\label{eq:galerkin_RB}
    \tilde{\mat{Z}}\vec{Q} + \tilde{\vec{F}}(\vec{WQ})=\tilde{\vec{P}}_\text{ext},
\end{equation}
where
\begin{align*}
    &\tilde{\mat{Z}}= \mt{W}{}{T}\mat{Z}\mat{W},\\
    &\tilde{\vec{F}}=\mt{W}{}{T}\vec{F}(\vec{WQ}),\\
    &\tilde{\vec{P}}_{\text{ext}}=\mt{W}{}{T}\vt{P}{ext}{}.
\end{align*}

\begin{algorithm}[H]
\caption{Basis construction}
 \label{algoritm:basis}
  \begin{algorithmic}[1]
  \REQUIRE  $a$ excitation amplitudes \eqref{eq:external_forces}, external force vector $\vec{P}_{\text{ext,}i}$ \eqref{eq:p_ext_freq}, static solution $\vt{u}{0}{}$ \eqref{eq:NLstaticeq}, modal frequency of interest $\overline{\omega}$ \eqref{eq:mode_freq},   \ $\mt{Z}{MC}{}$ \eqref{eq:Z_diag_term}, linear dynamic solution $\vt{U}{\emph{l}}{}$ \eqref{eq:LinDyn_freq},  $m$ amplification factors \eqref{eq:amplif}, multi-harmonic mass matrix $\overline{\mat{M}}$ \eqref{eq:Mmatrix}
  \ENSURE reduction basis $\mat{W} \in \mathbb{R}^{n(2H+1) \times p}$\\
  \emph{construct basis vectors}
  \FOR {$k=1:m$}
  \STATE $\vec{\theta}^{k} \leftarrow [\vt{u}{0}{T},\ \alpha_{\text{k}} \vt{U}{\emph{l}}{}(\overline{\omega})^{T}]$
  \STATE $\mat{J}^{k},\vec{F}^{k} \leftarrow$ \texttt{AFT}$^{1}$ $(\vec{\theta}^{\text{k}}, \overline{\omega})$
  
  \STATE $\vec{\gamma}_{1}^{k},\vec{\gamma}_{2}^{k} \leftarrow$ \texttt{EigenCond}$^{2}$ $(\overline{\mat{M}},\mat{J}^{k}, \overline{\omega})$
  
  \STATE $\mat{\Gamma}_\text{H}^{k} \leftarrow [\text{Real}\{{\vec{\gamma}_\text{1,H}^{k}}\},\text{Im}\{{\vec{\gamma}_\text{1,H}^{k}}\},\text{Real}\{{\vec{\gamma}_\text{2,H}^{k}}\},\text{Im}\{{\vec{\gamma}_\text{2,H}^{k}}\}]$
  \STATE $\mat{\Gamma}_{0}^{k} \leftarrow \texttt{StaticRet}^{3}(\mat{\Gamma}_\text{H}^{k}, \mat{J}^{k})$
  
  \STATE $\mat{\Gamma}^{k} \leftarrow
  \begin{bmatrix}
  \mat{\Gamma}_{0}^{k}\\ \mat{\Gamma}_\text{H}^{k}
  \end{bmatrix}$
  
  \FOR {$i=1:a$}
  \STATE $\vec{\overline{\gamma}}_\text{i}^{k} \leftarrow$ \texttt{AugForced}$^{4}$ $[\mt{Z}{MC}{},\mat{J}^{k},\vec{P}_{\text{ext,}i}]$
  
  \STATE $\mat{\Gamma}^{k} \leftarrow [\mat{\Gamma}^{k}, \vec{\overline{\gamma}}_\text{i}^{k}]$
  \ENDFOR 
  
  \emph{assign reduction basis of each harmonic}
  \STATE $\mt{V}{0}{} \leftarrow [\mt{V}{0}{}, \mat{\Gamma}_{0}^{k}]$\\
  \nonumber $\mt{V}{j}{c} \leftarrow [\mt{V}{j}{c}, \mat{\Gamma}_\text{j}^{k\text{,c}}]$\\
  \nonumber $\mt{V}{j}{s} \leftarrow [\mt{V}{j}{c}, \mat{\Gamma}_\text{j}^{k\text{,s}}]$\\
  \ENDFOR
  \STATE $\mt{V}{0}{} \leftarrow$ \texttt{SVD}$^{5}$ $(\mt{V}{0}{})$\\
  \nonumber $\mt{V}{j}{c} \leftarrow$ \texttt{SVD} $(\mt{V}{j}{c}$)\\
  \nonumber $\mt{V}{j}{s} \leftarrow$  \texttt{SVD} $(\mt{V}{j}{s})$
     \end{algorithmic} 
\end{algorithm}
\noindent
 $^{1}$\ \texttt{AFT} uses $\vec{\theta}^{k}$ as an input displacement to the AFT algorithm to compute the nonlinear forces $\vec{F}^{k}$ and the Jacobian of nonlinear forces $\mat{J}^{k}$.  \\
 $^{2}$\ \texttt{EigenCond} solves the eigenvalue problem \eqref{eq:eigen_lin} after applying the static condensation procedure using \eqref{eq:staticretrieval} and \eqref{eq:static_cond2}.\\
$^{3}$ \texttt{StaticRet} retrieves the static components of the eigenvectors \eqref{eq:staticretrieval}.\\
$^{4}$ \texttt{AugForced} computed the forced response of the $k$-th linear system according to \eqref{eq:augment}.\\
$^{5}$ \texttt{SVD} performs the singular value decomposition on the basis of each harmonic component.

\subsection{Hyper Reduction by Energy Conserving, Sampling, and Weighing (HR ECSW)}
\label{subsection:hyper_reduction_original}
Even though the size of the Augmented JP ROM in \eqref{eq:galerkin_RB} is significantly smaller than that of the original system, the speedups remain limited due to the high density of the mesh of nonlinear elements across the interface and the computational cost associated to the high number of sampling points required in the AFT scheme. Hyper Reduction (HR) refers to techniques developed to overcome the computational burden associated with the evaluation of the vector of nonlinear forces. In this section, we briefly review the Energy Conserving Mesh Sampling and Weighting (ECSW) method \cite{Remacle2012} method, and explain why we find it suitable for jointed structures. For these purposes, let us consider a data-driven ROM constructed in the time-domain. Using HF simulations, a reduction basis $\mat{S} \in \mathbb{R}^{n \times r}$, which approximates the displacements as

\begin{equation}
\vec{u}\text{(t)} \approx \mat{S} \vec{q}\text{(t)},
\label{eq:ROM_timedomain}
\end{equation}
can be constructed using, for instance, Proper Orthogonal Decomposition (POD), where $\vec{q}\text{(t)}$ is the vector of reduced coordinates. Subsequently, a Galerkin projection of the system onto $\mat{S}$ can be carried out. Since the dynamics of joints are known to be remarkably dissipative, it would be desirable to still preserve the dissipation of the system upon the approximation of the nonlinear forces. The dissipation over one time period due to frictional contact is

\begin{equation}
    \int_{\text{t}}^{\text{t+T}} \vt{f(t)}{}{T} \vec{\dot{u}(t)}\ \text{dt} \approx \int_{\text{t}}^{\text{t+T}} \vt{f(t)}{}{T} \mat{S} \vec{\dot{q}(\text{t})}\ \text{dt} =  \int_{\text{t}}^{\text{t+T}} \vt{\dot{q}(\text{t})}{}{T} \mt{S}{}{T}  \vec{f(t)}\ \text{dt}.
\end{equation}
Accurately approximating the projection of the nonlinear forces $\mt{S}{}{T}  \vec{f(t)}$ would therefore mean preserving the dissipation computed by the ROM. Indeed, this is what ECSW aims at, as it seeks an energetically equivalent approximation of the projection of the nonlinear forces $\tilde{\vecf{f}}\text{(t)}=\mt{S}{}{T}\vecf{f}\text{(t)}$, such that

\begin{equation}
\label{eq:idea_ECSW}
\tilde{\vecf{f}} = \sum_{e=1}^{\st{n}{e}{}} \mt{S}{e}{T} \vt{f}{e}{}(\mt{S}{e}{}\vt{q}{}{}) \approx \sum_{e \in E}  \xi_{\text{e}} \mt{S}{e}{T} \vt{f}{e}{}(\mt{S}{e}{}\vt{q}{}{}) = \tilde{\vecf{f}}_\text{HR},
\end{equation}
where $\st{n}{e}{}$ is the total number of nonlinear elements, and E represents only a small subset of elements for which the weights $\xi_{\text{e}}$ satisfy the condition $\xi_{e\in E} >0$. The training required to compute the vector $\vec{\xi} \in \mathbb{R}^{n_{e}}$, which consists of the weights of all nonlinear elements, is based on $n_s$ force snapshots $\begin{bmatrix} \vec{f}^{(i)}\end{bmatrix}_{i=1}^{n_s}$. Referring to the original method, these are snapshots of the nonlinear forces that are extracted from a HF time integration scheme. To compute the weights, a Non-Negative Least Squares (NNLS) optimization problem is set as
\begin{equation}
    \vec{\xi}^{*} = \text{argmin}_{\vec{\xi} \in \mathbb{R}^{n_{e}}, \vec{\xi} \geq 0} \norm{\mat{G}\vec{\xi} -\vec{b}}_{2}^{2},
    \label{eq:NNLS}
\end{equation}
where 
\begin{equation}
\mat{G} = \left. \begin{bmatrix}
        {\vec{g}}_{1,1} & \dots &  \vec{g}_{1,n_e} \\
    \vdots &  \ddots  & \vdots\\
    \vec{g}_{n_{s},1} & \dots  & \vec{g}_{n_{s},\st{n}{e}{}}\\
     \end{bmatrix} \right\} \text{$n_{s}$ snapshots,}
\end{equation}
with
\begin{equation}
\begin{split}
    \mt{g}{i,e}{} =  \mt{S}{e}{T} \vt{f}{e}{}(\mt{S}{e}{} \vec{q}^{\text{i}}),\\
    \vt{b}{i}{} =  \sum_{\text{e}}^{\text{n}_\text{e}} {\mt{g}{i,e}{} },
    \end{split}
\end{equation}
where $\mat{G} \in \mathbb{R}^{r (n_{s} ) \times n_e}$, $\vec{b} \in \mathbb{R}^{r (n_{s}) \times 1}$, $\vt{g}{i,e}{} \in \mathbb{R}^{r \times 1}$, and $\vt{b}{i}{}\in \mathbb{R}^{r \times 1}$. However, the optimal solution to \eqref{eq:NNLS} is NP-hard. Therefore, a sub-optimal, yet feasible, algorithm that is known as the sparse NNLS \cite{Peharz2012}, which terminates the search for a solution once the objective function returns a value below a certain threshold, is used. This threshold is controlled by a user-set tolerance $\tau$, and is given by

\begin{equation}
\label{eq:nnls}
   \norm{\mat{G}\vec{\xi} -\vec{b}}_2 \leq \tau  \norm{\vec{b}}_2.
\end{equation}

\subsection{Training for Frictional Contact in Frequency Domain}
\label{subsec:hyper_reduction_training}

\begin{figure}
\centering
\includegraphics[scale=0.30]{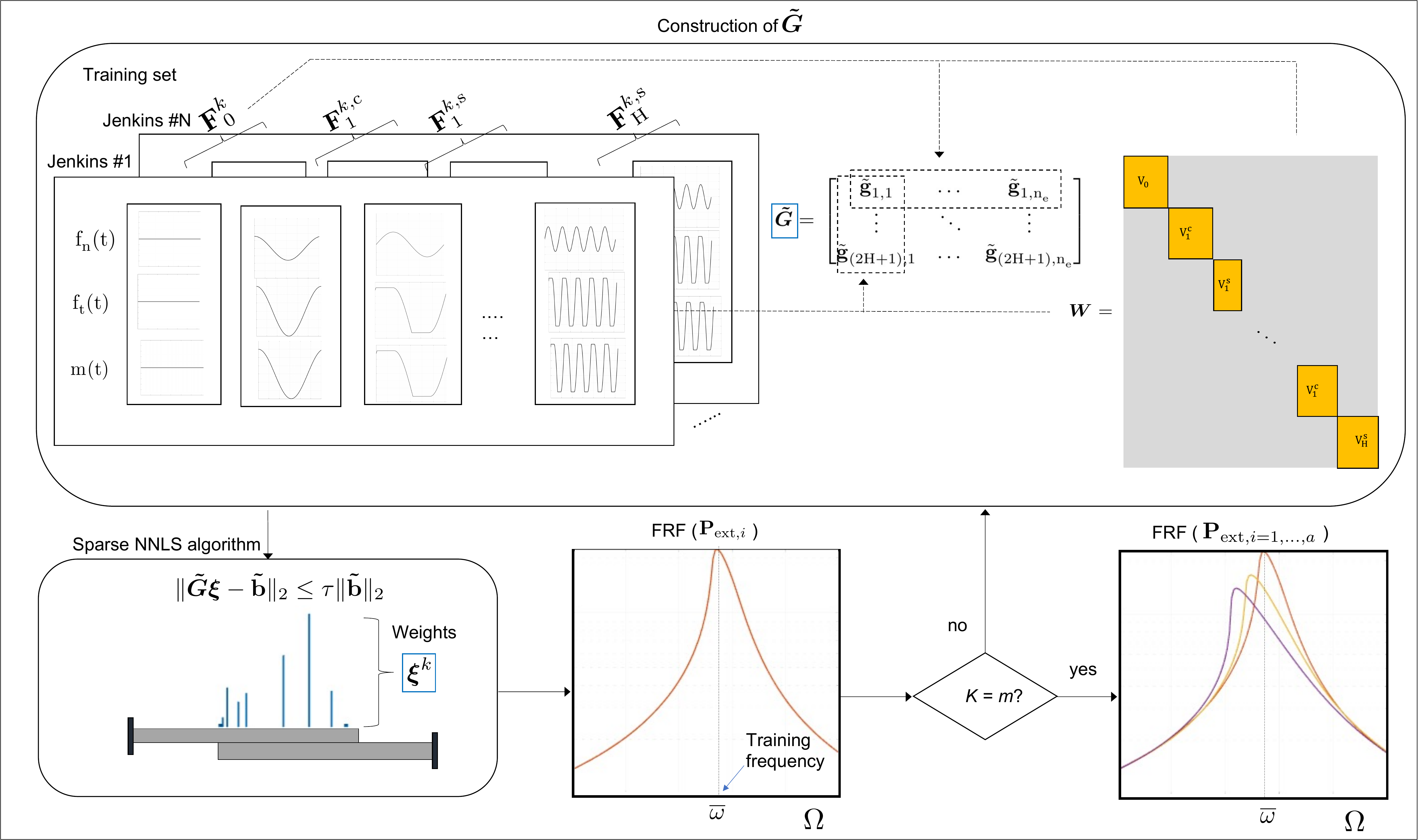}
\caption{Flowchart of the method 2/2 - Hyper Reduction determined at $\Omega = \overline{\omega}$. The same hyper mesh and weights are used then across the frequency sweep.}
\label{fig:flowchart_HR}
\end{figure}

We propose HR training that does not rely on HF simulations, and is in line with frequency domain analysis in MHBM. Intuitively, the information conveyed in a set of multiple time snapshots of the solutions can also be contained in the harmonic components of the nonlinear force vector. Therefore, for a given amplitude of excitation, we use the harmonic components of $\vec{F}^{k}$, which is an output of the the AFT in step 2 of Algorithm \ref{algoritm:basis}, as our 2H+1 force snapshots. This means that this training set does not incur any additional costs after the construction of the reduction basis. Considering the reduction basis $\mat{W}$ and the Augmented JP ROM \eqref{eq:galerkin_RB}, the approximation we seek becomes
\begin{equation}
\label{eq:idea_ECSW}
\tilde{\vecf{F}} = \sum_{e=1}^{\st{n}{e}{}} \mt{W}{e}{T} \vt{F}{e}{}(\mt{W}{e}{}\vt{Q}{}{}) \approx \sum_{e \in E}  \xi_{\text{e}} \mt{W}{e}{T} \vt{F}{e}{}(\mt{W}{e}{}\vt{Q}{}{}) = \tilde{\vecf{F}}_\text{HR}.
\end{equation}
The same sparse NNLS algorithm \cite{Peharz2012} is used to determine $\vec{\xi}$. However, we define an alternative matrix $\mat{\tilde{G}}$ as
\begin{equation}
\mat{\tilde{G}} = \left. \begin{bmatrix}
        \tilde{\vec{g}}_{1,1} & \dots &  \tilde{\vec{g}}_{1,\st{n}{e}{}} \\
    \vdots &  \ddots  & \vdots\\
    \tilde{\vec{g}}_{\text{(2H+1),1}} & \dots  & \tilde{\vec{g}}_{\text{(2H+1)},\st{n}{e}{}}\\
     \end{bmatrix} \right\} (\text{$2H+1$) snapshots,}
\end{equation}
where 
\begin{equation}
\begin{split}
    \tilde{\mat{g}}_{\text{1,e}}  &= [\mt{V}{0,e}{}]^{\text{T}}[\vt{F}{e}{}]_{0}, \\ \tilde{\mat{g}}_{\text{2,e}}  &= [\mt{V}{1,e}{c}]^{\text{T}}[\vt{F}{e}{}]_{1}^{c}, \\
    \tilde{\mat{g}}_{\text{3,e}}  &= [\mt{V}{1,e}{s}]^{\text{T}}[\vt{F}{e}{}]_{1}^{s},
\label{g_training_HY}
\end{split}
\end{equation}
and $\vt{\tilde{b}}{i}{}$ then becomes
\begin{equation} 
    \vec{\tilde{b}}_{\text{i}} =  \sum_{\text{e}=1}^{\st{n}{e}{}} {\mat{\tilde{g}}_{\text{i,e}}} .
\end{equation}
Regarding the value of $\tau$, a value of 0.01 has been found to lead to sparse hyper meshes capable of producing accurate results. Figure \ref{fig:flowchart_HR} shows a flowchart of the hyper reduction step.

\subsection{Error Metric}
\label{subsec:error_metric}
Since we aim to construct reliable ROMs which no longer require HF simulations, we propose an error metric on the ROM solution that does not rely on a HF reference solution. Since approximating the displacements using \eqref{eq:QAnsatz} and \eqref{eq:reduced_disp} leads to residual forces $\vec{r}(\text{t})$, where

\begin{equation}
    \vec{r}(\text{t}) = \mat{M}\vec{\ddot{u}_{h}}(\text{t}) + \mat{C}\vec{\dot{u}_{h}}(\text{t}) + \mat{K}\vt{u}{h}{}(\text{t}) + \vec{f}  (\vt{u}{h}{}(\text{t}),\vt{\dot{u}}{h}{}(\text{t})) - \vec{P}_{\text{ext,}i}(\text{t}),
    \label{eqn:residual}
\end{equation}
we therefore evaluate the error $E$ on the residual forces at the DOF of excitation (subscript y) using a root mean square (RMS) measure as

\begin{equation}
    E=\frac{\sqrt{\frac{1}{N} \sum\limits_{\text{i}}^{N}[\st{r}{y}{}(\st{t}{i}{})]^{2}}}{\sqrt{\frac{1}{N}\sum\limits_{\text{i}}^{N}  [\st{p}{E,y}{}(\st{t}{i}{})]^{2}}},
    \label{eq:error_metric}
\end{equation}
where $N$ is the number of sampling points in the time domain. It will be shown in Section \ref{sec:case_study_beam} and \ref{sec:case_study_frame}, that for the test cases we studied, errors below  15\% indicate a HR-ROM solution with an accuracy capable of capturing the global amplitude-dependent stiffness and damping of the system.

\section{Case Study: Jointed Beam}
\label{sec:case_study_beam}
\subsection{FE Model}
\begin{figure} [h!]
\centering
\includegraphics[scale=0.13]{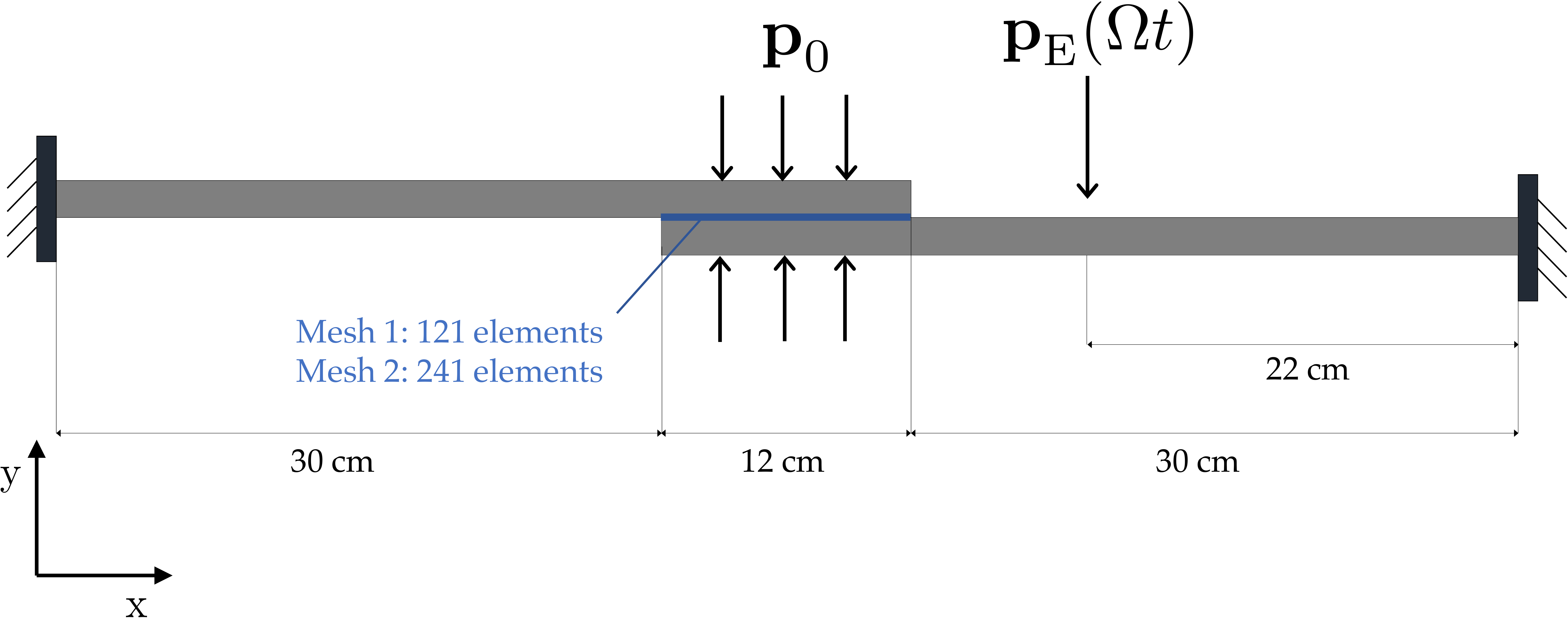}
\caption{Jointed beam - Test case 1}
\label{fig:TestBeam}
\end{figure}

Figure \ref{fig:TestBeam} shows a prototype structure of the Brake-Reu{\ss} beam \cite{Brake2018b}. The structure is composed of two beams joined together by the preclamping action of 3 bolts. Each beam has dimensions 42x2.5x2.5 cm. The material is steel with Young's Modulus E = 189 GPa and density $\rho$ =7820 Kg/$\text{m}^{3}$. The beams are modelled using planar Euler-Bernoulli beam elements. The contact interface is 12 cm long, and the frictional contact is modelled using an array of node-to-node Jenkins elements with unilateral springs. The contact parameters are assumed as $\text{k}_{\text{t}} = 3 \times 10^{11}\ \text{N/m}^{2}$, $\text{k}_{\text{n}} = 4 \times 10^{12}\ \text{N/m}^{2}$, and $\mu$=0.4. Each bolt exerts a force of 1.25 KN, which is modelled as concentrated forces applied at nodes lying within the tributary area of each bolt. The material damping is modelled as Rayleigh damping with a modal damping ratio  $\zeta = 3 \times 10^{-3}$. The excitation force acts in the y-direction, with amplitudes $\st{A}{}{} = 0.1, 2,5,10$ N, frequencies $\Omega$ ranging from 205 to 230 Hz, and is positioned at a distance 22 cm from the right end. We assume 5 harmonics in the Ansatz. The analysis is carried out using an in-house MATLAB code and implements sequential continuation\footnote{An attempt to apply a numerical continuation method (e.g. arc-length parameterization) for the HR-ROM did not lead to converging results. Unlikely to be a limitation, since the same procedure is problem-free for the next test case we will present, we report this experience as an anomaly. It should also be kept in mind that a failure in convergence is a flag to the analyst, as opposed to an inaccurate result, which if present in a converged solution, would be detected by the error metric. Lastly, the application of arc-length continuation for the HF simulation of the beam test case did not show any new insights compared to sequential continuation, therefore the presented results have been concluded valid.}. We monitor the ratio between the maximum displacement at the DOF of excitation and the amplitude of excitation.

\noindent
We perform our analysis using two different meshes, namely: 
\begin{itemize}
    \item Mesh (1): 121 Jenkins elements for the contact interface and a total of 300 beam elements.
    \item Mesh (2): 241 Jenkins elements for the contact interface and a total of 540 beam elements. 
\end{itemize}

\subsection{Results}
\begin{figure}[hbt!]
    \centering
    \begin{minipage}{0.5\textwidth}
        \centering
        \includegraphics[width=1\textwidth]{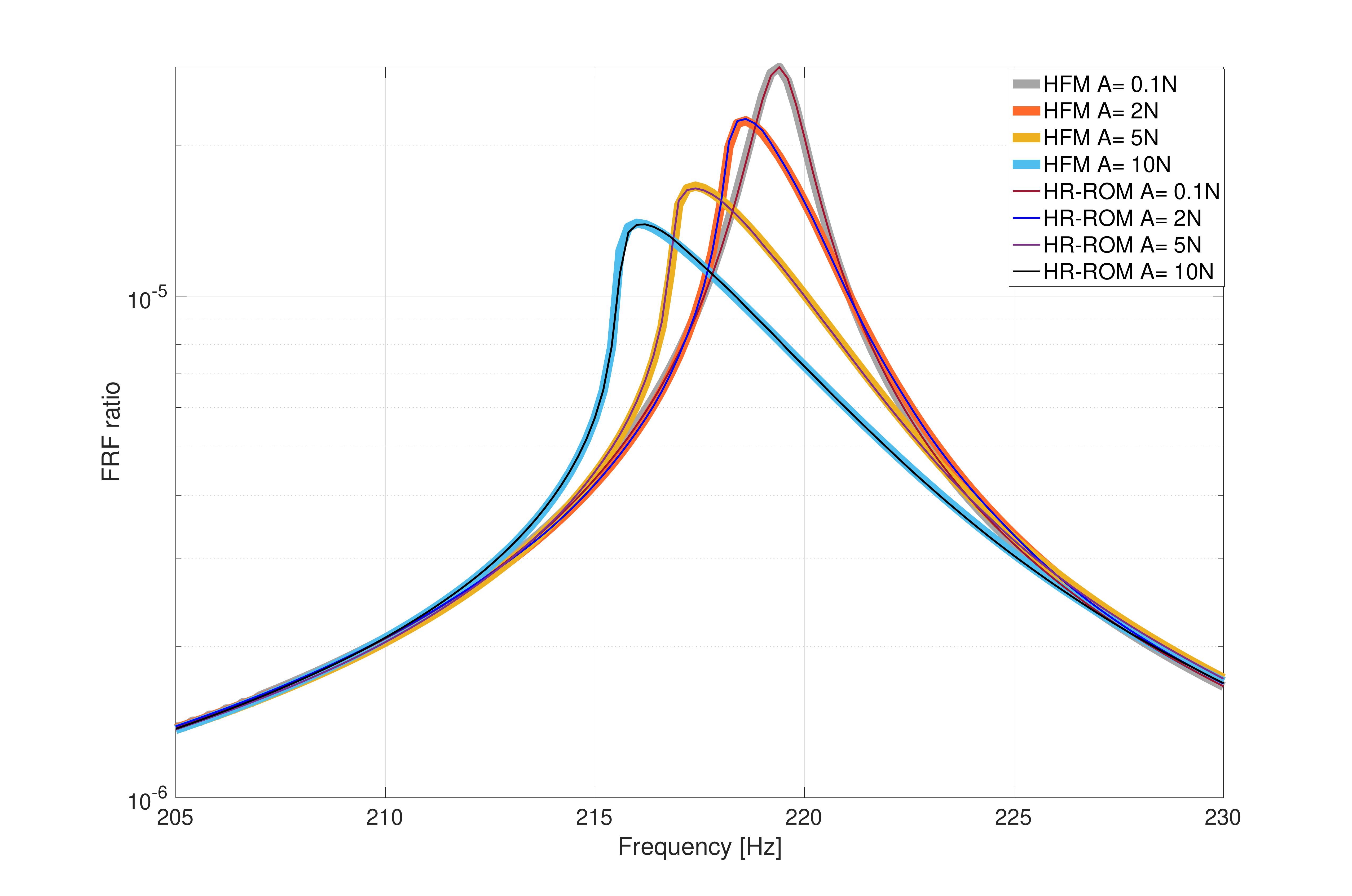} 
        \caption{FRF for Mesh (1) using HF and HR-ROM}
        \label{fig:FRF_beam_Mesh1}
    \end{minipage}\hfill
    \begin{minipage}{0.5\textwidth}
        \centering
        \includegraphics[width=1\textwidth]{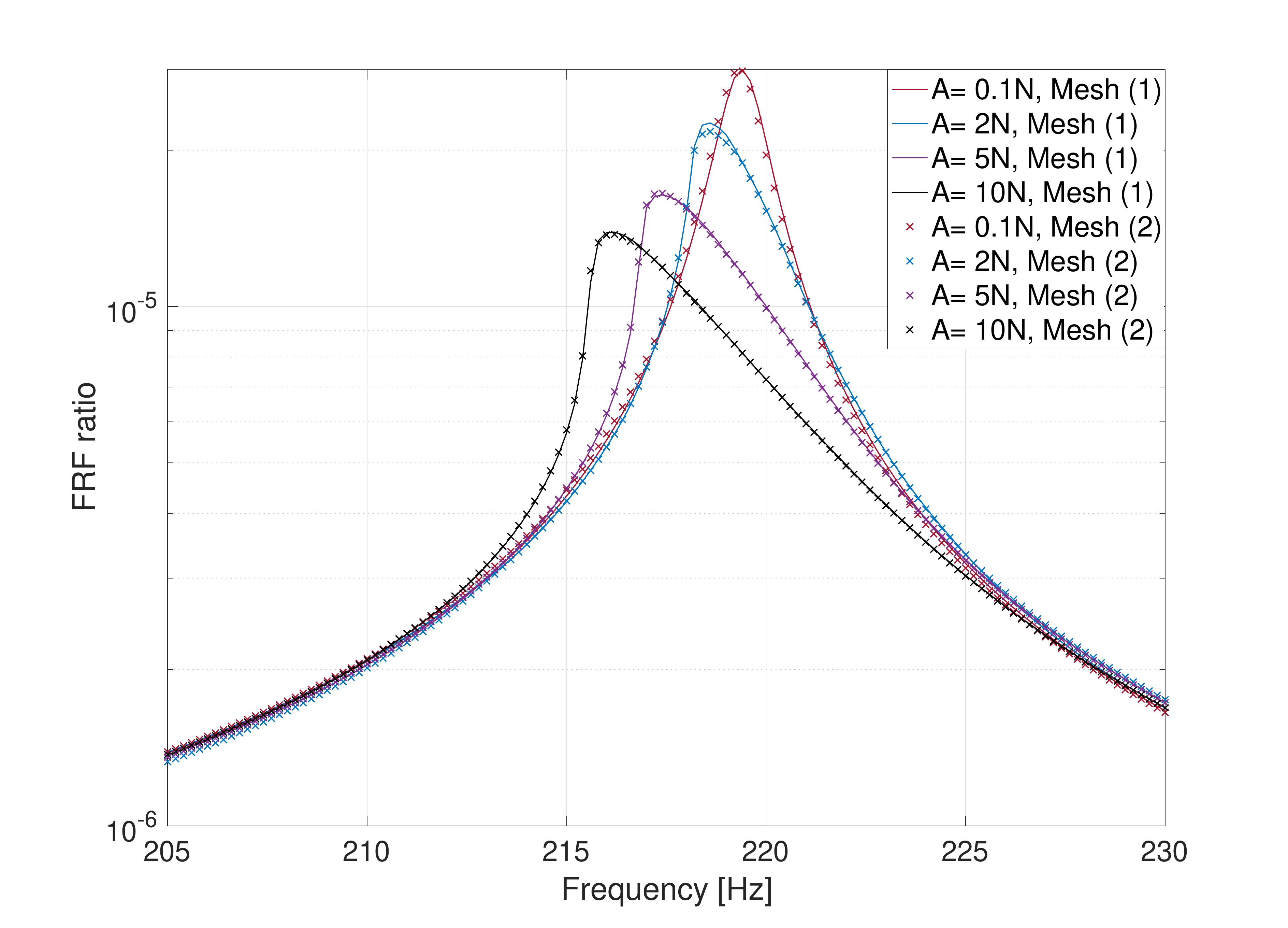} 
        \caption{FRFs for Mesh (1) and Mesh (2) using HR-ROM}
        \label{fig:FRF_beam_HR}
    \end{minipage}
\end{figure}

\begin{figure} [h]
\centering
\includegraphics[height=4cm, width=18cm]{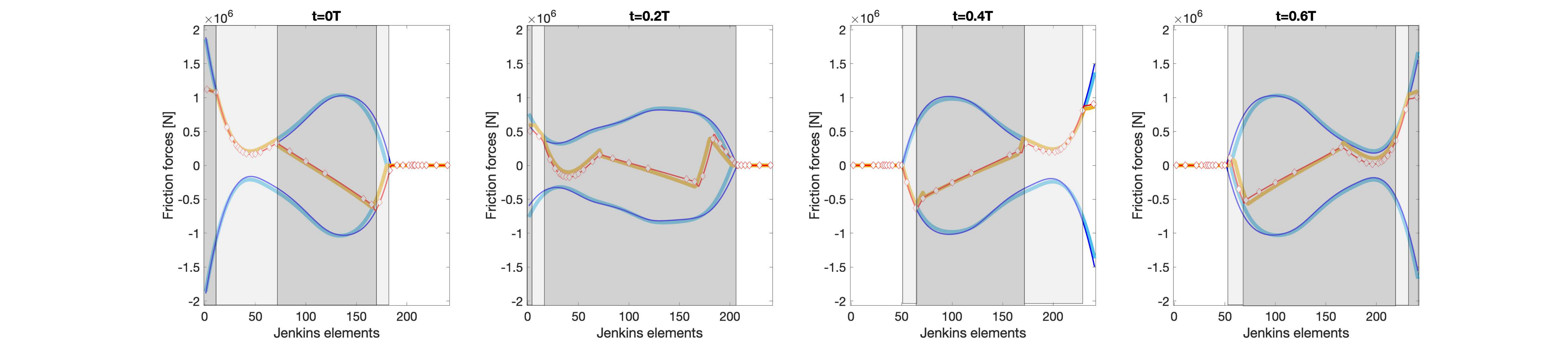}
\caption{Snapshots of friction forces along the interface for Mesh (1) across the time period T due to $\text{A}=\ 10\text{N}$, $\Omega=\ 216.2\text{Hz}$}.
\label{fig:BeamFriction_snaps}
\end{figure}

Figure \ref{fig:FRF_beam_Mesh1} shows the FRFs for Mesh (1) around the first bending mode using the HFM (9900 DOFs), and the HR-ROM (75 DOFs). As shown, the results of HR-ROM closely match those of the HFM, and the online speedups range from 5.5 to 7.0, as shown in Table \ref{table:beam_online_costs}. The offline costs reported in Table \ref{table:beam_offline_costs}. Note that the basis is constructed only once regardless of the number of analysis performed, and that the total time for the training for hyper reduction is less than 1 second. As for the hyper-mesh, Table \ref{table:beam_hyper_mesh} shows that the ratio of hyper-elements to total number of Jenkins elements ranges from 12\% to 24\%. Using the error metric, the maximum error around resonance for A=10 N is less than 9\%. Since the error metric considers residual forces, and it is known that the convergence on the forces is slow compared to the displacements \cite{Krack2017}, one should in general expect the presence of errors even in the case of overlapping FRFs. Nonetheless, the ROM has the capacity for reproducing the evolution of friction forces with good accuracy. For instance, the distributions of friction forces at different time instances for A= 10N and $\Omega$= 216.2 Hz, according to the HFM and HR-ROM, respectively, are plotted in Figure \ref{fig:BeamFriction_snaps}. This case corresponds to an error of 3.25\% for the HR-ROM; nevertheless, the agreement is good.

\begin{figure}[hbt!]
\centering
\includegraphics[scale=0.2]{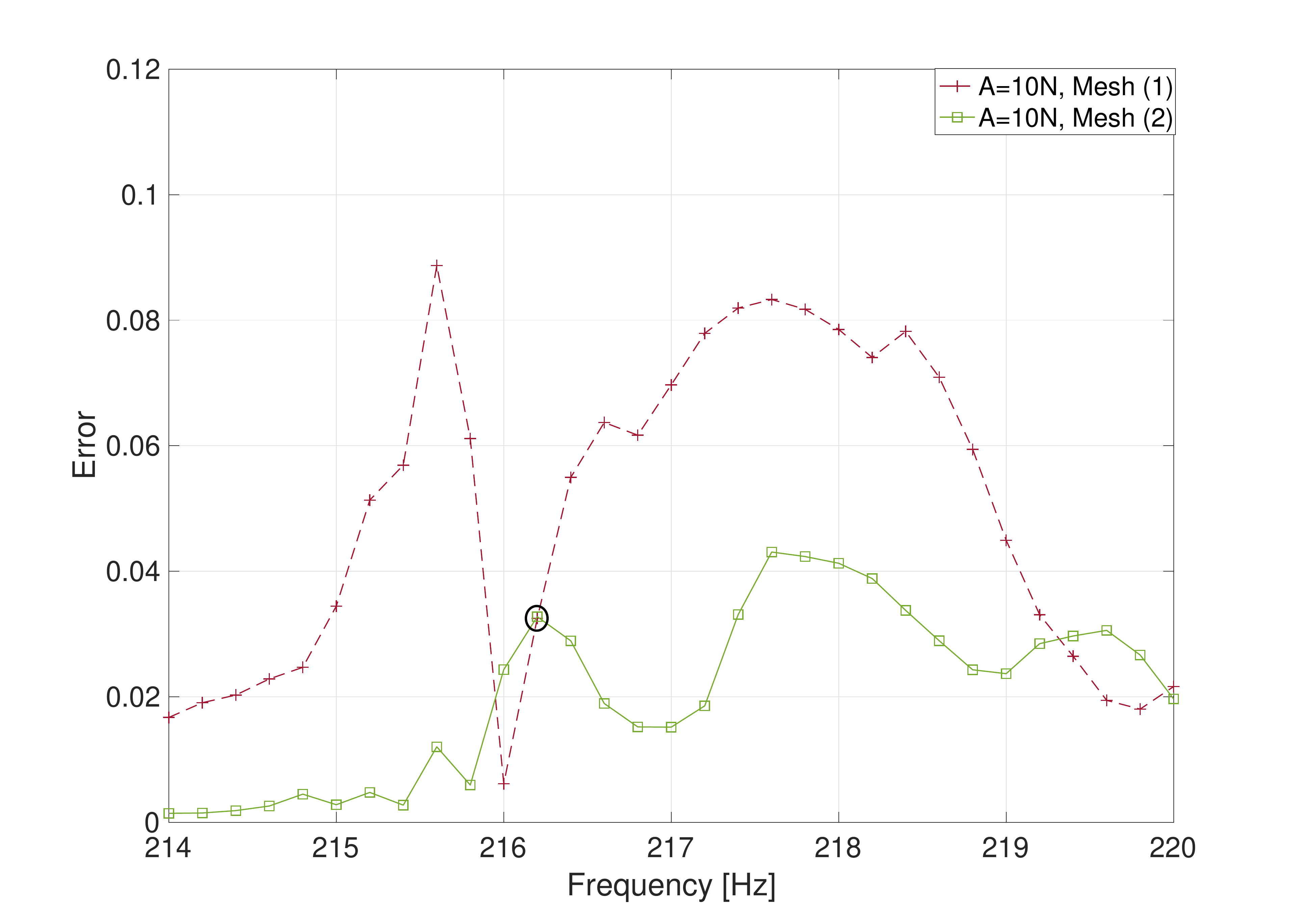}
\caption{Errors around resonance for A= 10N using HR-ROM for Mesh (1) and Mesh (2). The red square locates an error of 3.25\%, for which the corresponding response along the contact interface is plotted in Figure \ref{fig:BeamFriction_snaps}.
\label{fig:beam_errors}}
\end{figure}

To test the convergence of the model with respect to the number of nonlinear elements, we analyze the structure using Mesh (2). The total number of reduced DOFs increases only marginally, from 75 to 77. Figure \ref{fig:FRF_beam_HR} indicates that Mesh (1) is already convergent, whereas Table \ref{table:beam_online_costs} shows speedups that now range from 9.5 to 42.8. Noting the significant increase in speedups for the finer mesh, we highlight another advantage of the method: the computational burden of simulations aimed to determine an appropriate mesh size for a structure, whose response is unforeseen, is largely alleviated. Indeed, although the nonlinear elements in Mesh (2) are twice as much as in Mesh (1), the hyper mesh only increases by around 10\% for A= 10N, for example, as shown in Table \ref{table:beam_hyper_mesh}. In practical situations, it is not trivial to determine apriori the right mesh size for a desired accuracy, and often tedious mesh convergence studies need to be carried out. The example discussed here shows that doubling the number of nonlinear elements results in a tenfold more expensive HFM. The HR-ROM counterpart, however, features a relatively marginal increase in solution time from 95.1s to 129.7s. This is mainly due to hyper-reduction, which selects just enough elements to represent the nonlinearity, irrespective of the underlying mesh density.

\begin{table}[hbt!]
\centering
\begin{tabular}{c|ccc|ccc|c}
\toprule

              & \multicolumn{3}{c|}{Mesh (1)}                                                & \multicolumn{3}{c|}{Mesh   (2)}                                              &  \\ 
\midrule
Force [N] & HR [s] & HF [s] & Speedups & HR [s] & HF[s] & Speedups & Ratio of Speedups     \\ \midrule
0.1         & 29.5         & 206.4      & 7.0      &  57.7       & 550.7      & 9.5     & 1.4                   \\
2           & 68.7       & 404.2      & 5.9      & 100.9       & 1864       & 18.5     & 3.1                   \\
5           & 85.7       & 473.5      & 5.5      & 115.7      & 3217.6     & 27.8     & 5.0                   \\
10          & 95.1       & 520.4      & 5.5      & 129.7      & 5547.8     & 42.8     & 7.8      \\  \bottomrule     
\end{tabular}
\caption{HF and ROM run times for Mesh (1) and Mesh (2)}
\label{table:beam_online_costs}
\end{table}

\begin{table}[h]
\parbox{.45\linewidth}{
\centering
\begin{tabular}{c|cc}
\toprule
Offline   cost                                    & Mesh (1) & Mesh (2) \\
\midrule
Total training   time [s] & 0.35     & 0.86      \\
Basis time {[}s{]}                                & 28.4     & 97.5    \\ \bottomrule 
\end{tabular}
\caption{ROM offline costs for Mesh (1) and Mesh (2)}
\label{table:beam_offline_costs}
}
\hfill
\parbox{.45\linewidth}{
\centering
\begin{tabular}{c|cc}
\toprule
F [N] & Hyper Mesh (1) & Hyper Mesh (2) \\
\midrule
0.1       & 14             & 19             \\
2         & 23             & 26             \\
5         & 27             & 29             \\
10        & 29             & 32         \\ \bottomrule   
\end{tabular}
\caption{Hyper mesh for Mesh (1) and Mesh (2)}
\label{table:beam_hyper_mesh}
}
\end{table}

\section{Case Study: Frame}
\label{sec:case_study_frame}
\subsection{FE Model}
\begin{figure}
    \centering
    \begin{minipage}{0.65\textwidth}
        \centering
        \includegraphics[scale=0.1]{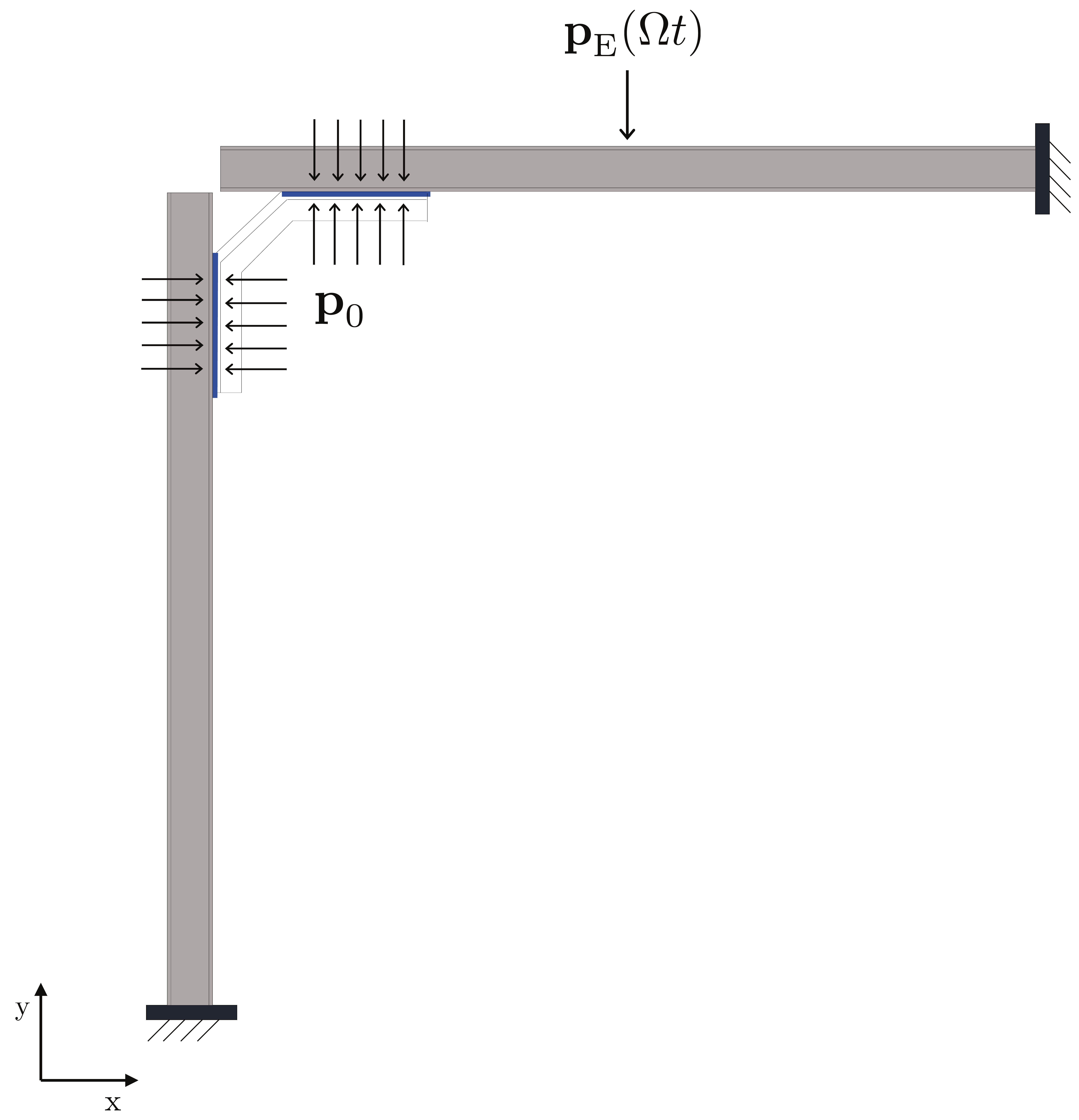}
        \caption{Frame structure- Test case 2}
\label{fig:frame_sketch}
    \end{minipage}\hfill
    \begin{minipage}{0.35\textwidth}
        \centering
        \includegraphics[width=1\textwidth]{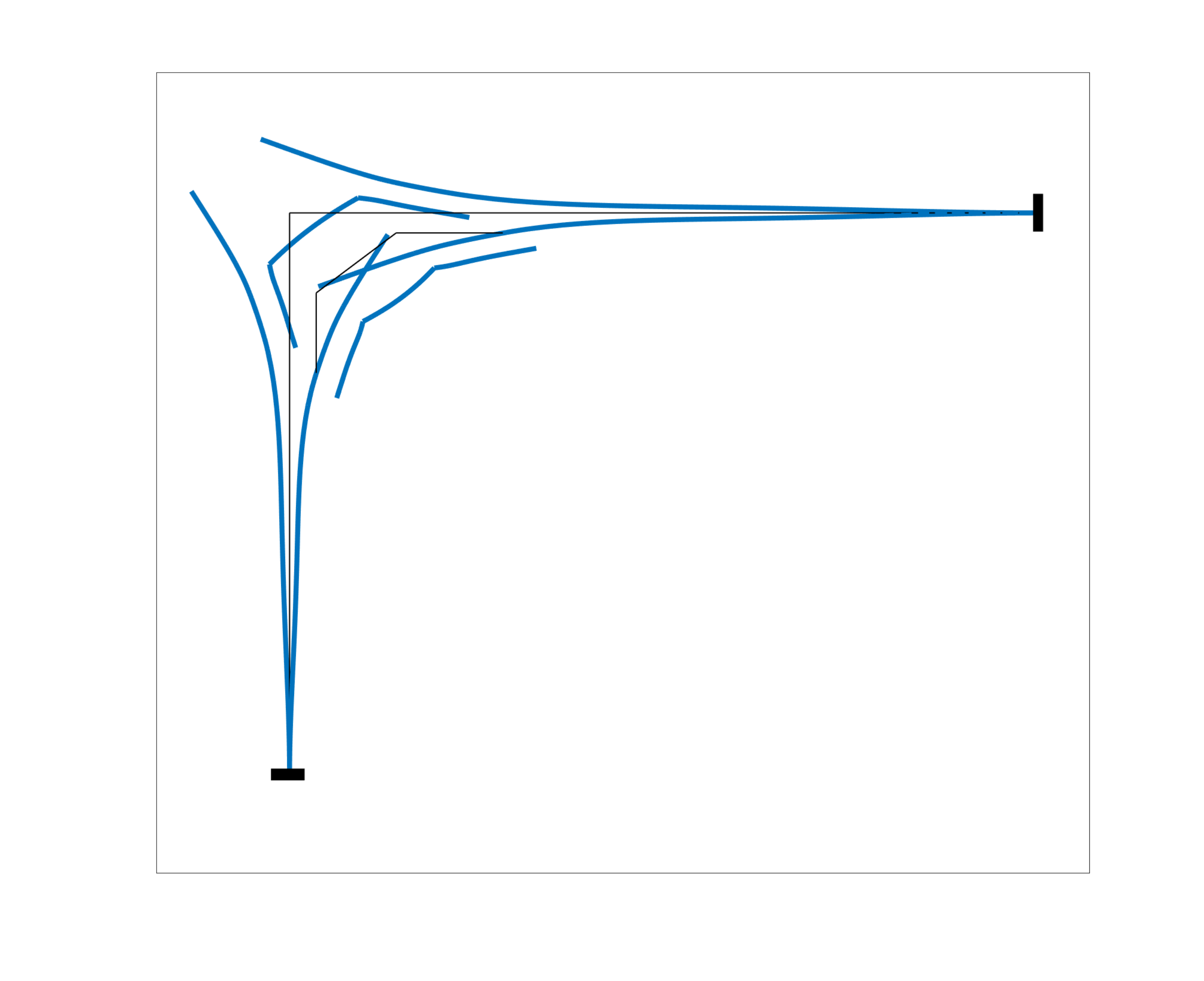} 
        \caption{$2^{\text{nd}}$ mode shape}
        \label{fig:frame_mode}
    \end{minipage}
\end{figure}

Figure \ref{fig:frame_sketch} shows a frame that is composed of a vertical column and a beam, both of length 2.80m, and made of a steel HEA 300 cross-section, with an area of $112\ \text{cm}^{2}$, and a second moment of area of $18260\  \text{cm}^{4}$. The column and the beam are joined by a bolted steel bracket having a T-cross section, whose area is $37.5\ \text{cm}^{2}$ and second moment of area of $3652\ \text{cm}^{4}$. The structure is modeled using planar Euler-Bernoulli beam elements. Each contact interface is 40 cm long, is subject to the preclamp action of 5 bolts, and discretized using 121 Jenkins elements. The FE discretization results in 1743 spatial DOFs.  Each bolt exerts a force of 6.75 tons, which is modelled in terms of concentrated forces applied at the nodes existing within the tributary area of each bolt. The contact parameters are assumed as $\st{k}{t}{}=3e11\ \text{N/m}^{2}$, $\st{k}{n}{}=4e12\ \text{N/m}^{2}$, and $\mu=0.3$. The excitation force is applied at a distance 1.4m from the fixed end, with amplitudes ranging from 100N to 1500N. The excitation frequency targets the second mode shape of the linearized structure, which is displayed in Figure \ref{fig:frame_mode}. The modal damping ratio is 3e-3. Numerical continuation is used to solve the MHBM system of equations, using NLvib \cite{Krack2019}. The continuation uses tangent predictor and arc-length corrector steps. Concerning the number of harmonics, we carried out a convergence study which showed that an Ansatz of 3 harmonics delivers satisfactory accuracy. The total number of DOFs for the MHBM equations becomes 12201. The HR tolerance $\tau$ is set to $10e-3$. Two reduction basis are considered for this study, namely, Basis (1) with 4 amplifications (i.e., $\alpha_{\text{k}=1,..,4}$ ), and Basis (2) with 6 amplifications. These are denoted in the tables and figures by B(1) and B(2).

\subsection{Results}
\begin{figure}
    \centering
    \begin{minipage}{0.5\textwidth}
        \centering
        \includegraphics[width=1\textwidth]{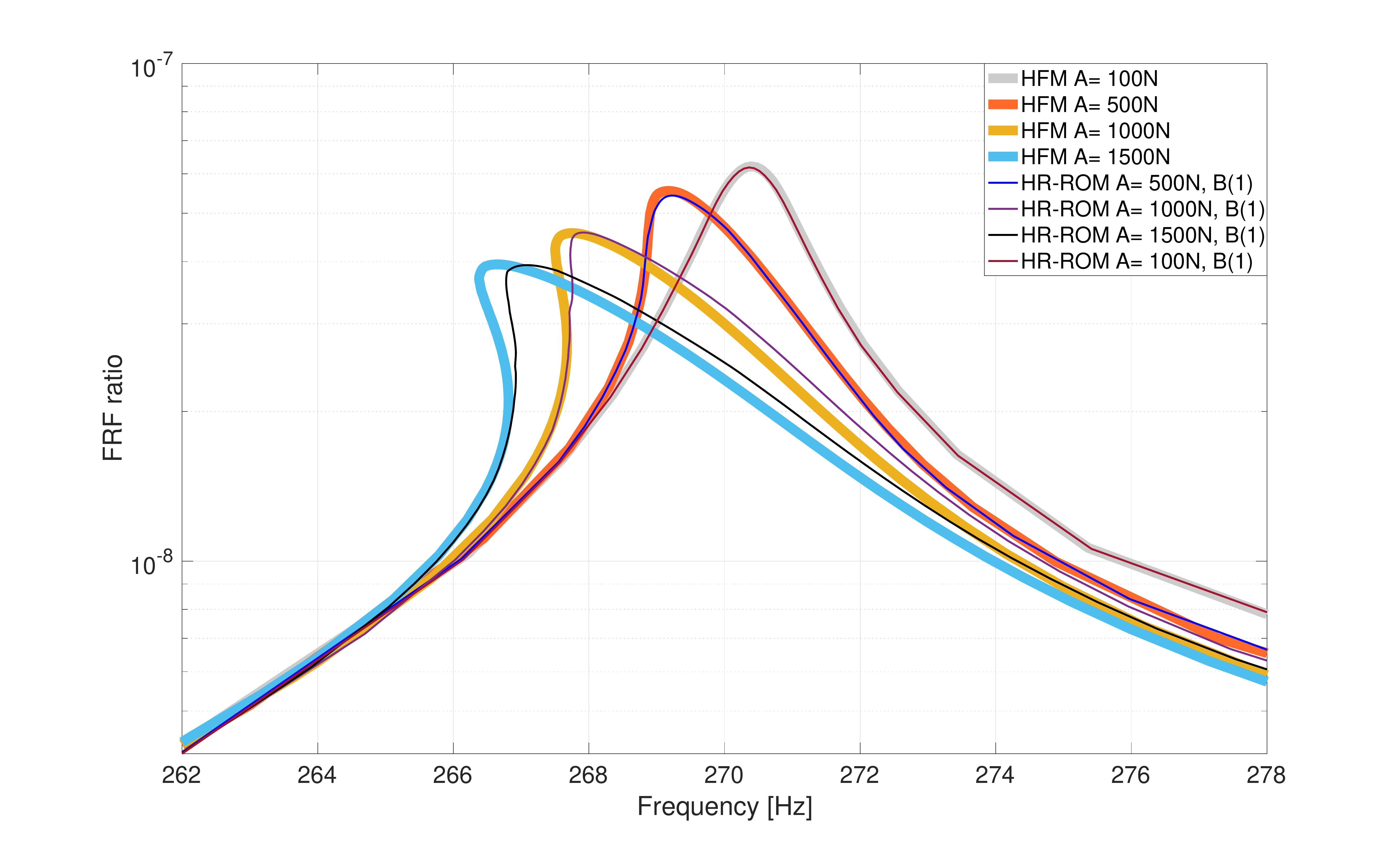} 
        \caption{FRFs for HF vs HR - Basis (1)}
        \label{fig:frame_frfs_basis1}
    \end{minipage}\hfill
    \begin{minipage}{0.5\textwidth}
        \centering
        \includegraphics[width=1\textwidth]{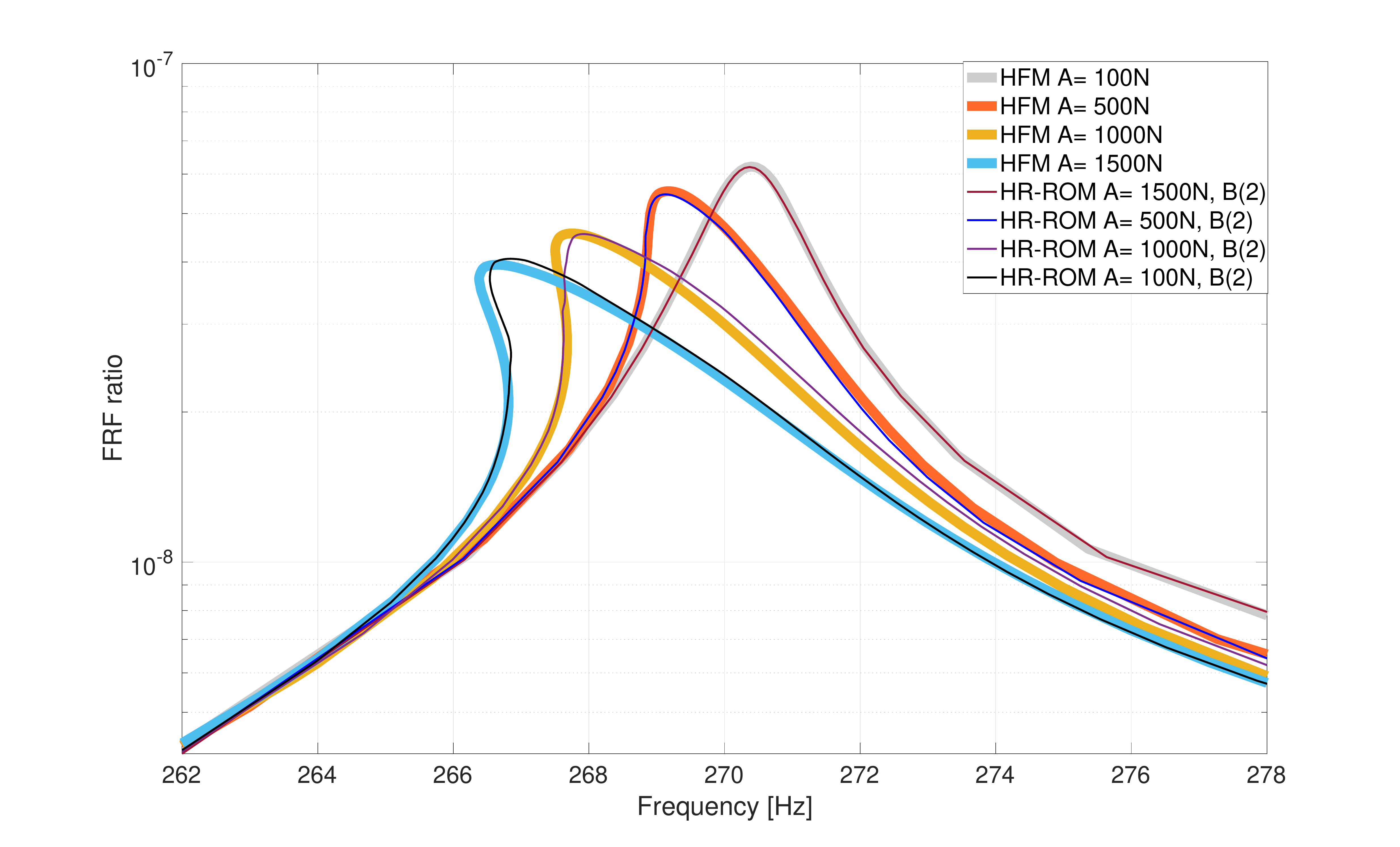} 
        \caption{FRFs for HF vs HR - Basis (2)}
        \label{fig:frame_frfs_basis2}
    \end{minipage}
\end{figure}

Figure \ref{fig:frame_frfs_basis1} shows the FRFs of the HFM and HR ROM using Basis (1). The total number of reduced DOFs is 90 in this case. Even though the agreement is good for low amplitudes, notable deviations of the responses, in particular, at A=1500N are shown. Since the ROM should be operable without resorting to HFM, one should be able to detect such inaccuracies through the error metric. The errors, shown in Figure \ref{fig:frame_errors}, reach 21\% around resonance. Furthermore, one should expect converging results upon the enrichment of the basis of the ROM. Consequently, two additional amplifications are included to construct the second reduction basis B(2). The total number of reduced DOFs increases from 90 to 110. As shown in Figure \ref{fig:frame_frfs_basis2}, the FRFs show significant improvement. Moreover, the overall errors  are shown in Figure \ref{fig:frame_errors} to decrease with the maximum error around resonance falling below 15\%. In fact, an error threshold of 15\% has repeatedly proved indicative of a reliable depiction of the amplitude-dependent dissipation and stiffness for the test cases we studied.

\begin{figure}[h!]
\centering
\subfloat[Column interface]{\includegraphics[height=4cm, width=18cm]{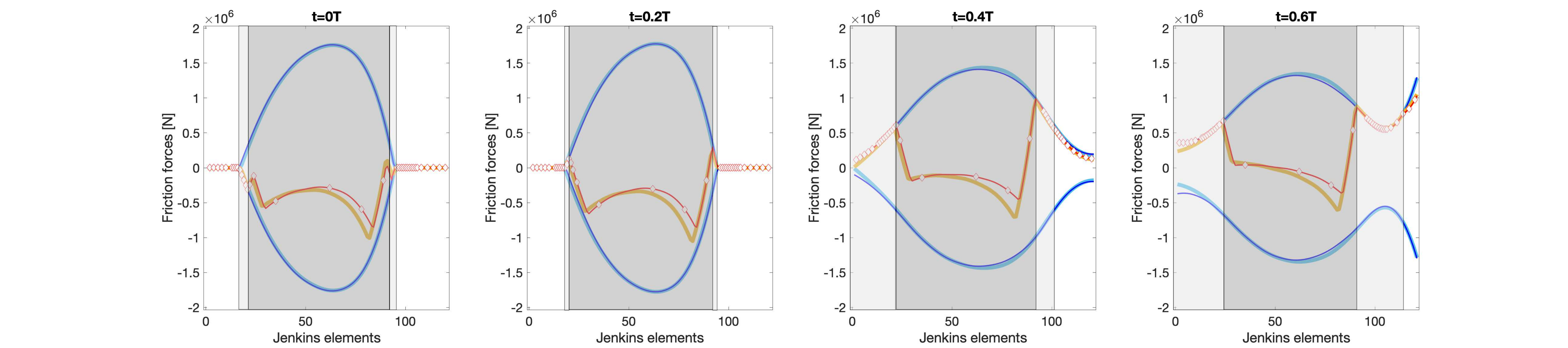}}\\
\subfloat[Beam interface]{\includegraphics[height=4cm, width=18cm]{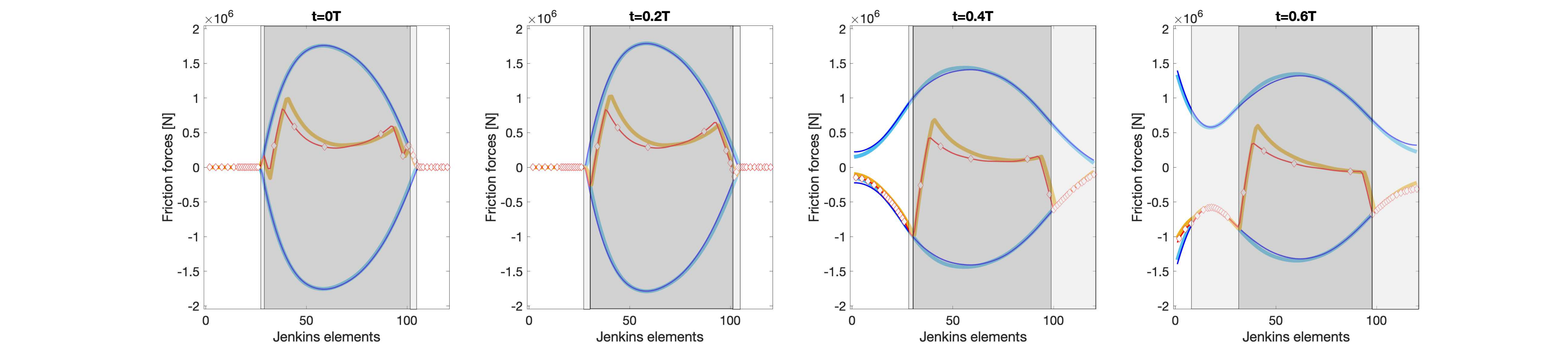}}\\
\subfloat{\includegraphics[width=13cm]{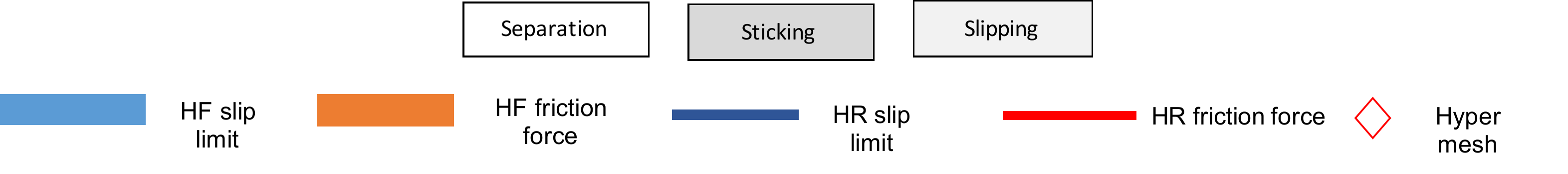}}
\caption{Time snapshots of friction forces along the contact interface of the column (top) and of the beam (bottom) for A= 1500N at $\Omega$= 267.83 Hz for HR-ROM (red rectangle in Figure \ref{fig:frame_errors}), and $\Omega$=267.88 Hz for HFM}
\label{fig:frame_snaps}
\end{figure}

\begin{figure}[h!]
\centering
\includegraphics[scale=0.2]{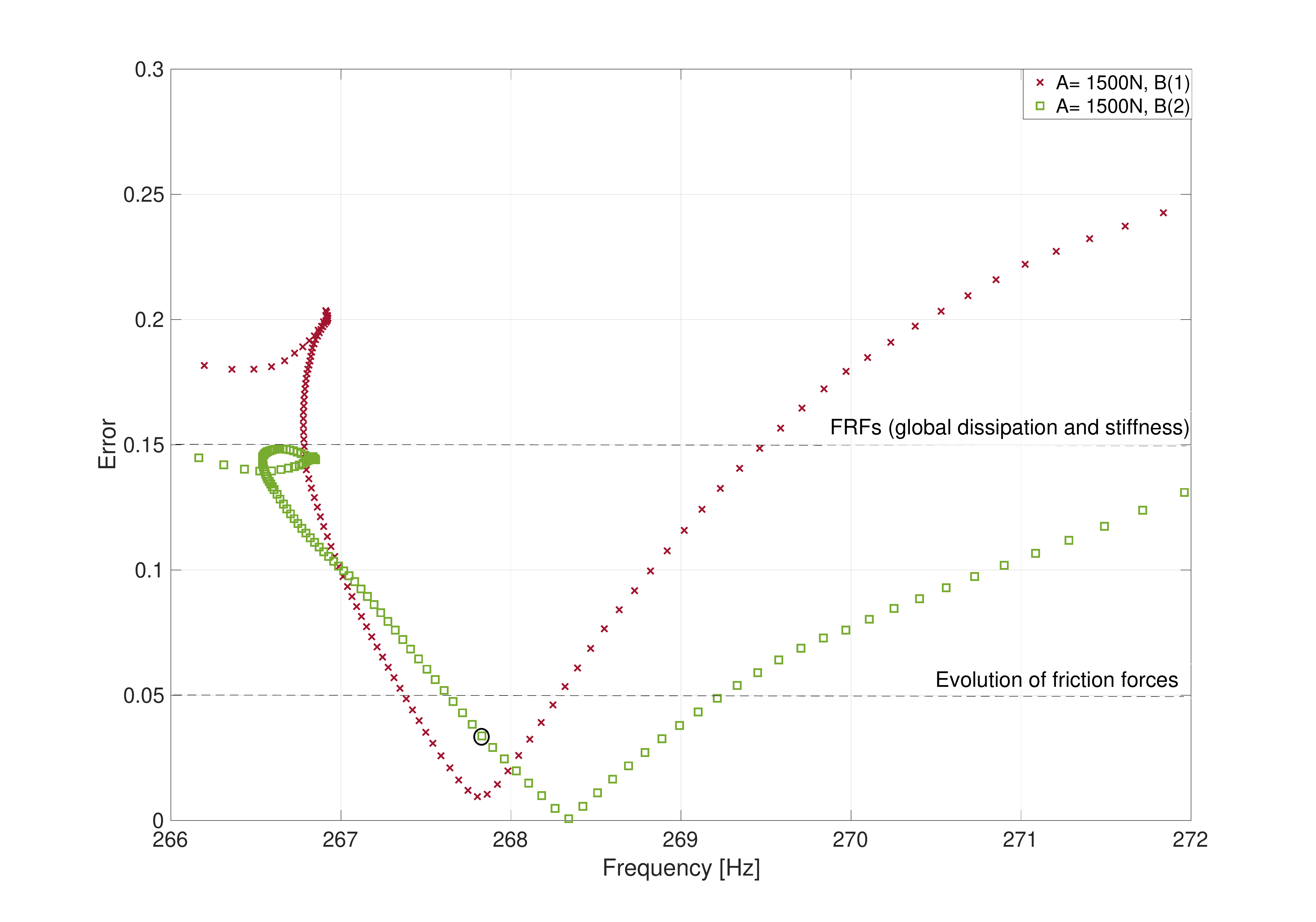}
\caption{Errors computed around resonance for A= 1500N using HR-ROM for Basis (1) and Basis (2). The black circle locates the error corresponding to $\Omega$=267.83 Hz whose corresponding response along the contact interface is plotted in Figure \ref{fig:frame_snaps}.}
\label{fig:frame_errors}
\end{figure}

As discussed in Test Case (1), the forces are expected to exhibit slower convergence. Accordingly, a lower threshold for an accurate spatial and temporal evolution of frictional contact forces was shown to be around 5\%. For instance, the time snapshots of friction forces for $\Omega \approx 267.8$  Hz are plotted in Figure \ref{fig:frame_snaps} along both interfaces. This is to say, that depending on the purpose of the analysis, the user can enrich and accordingly assess the ROM. The online and offline costs are reported in Tables \ref{table:frame_online_costs} \& \ref{table:frame_offline_costs}. The speedups using Basis (2) range from 11.9 to 18.5. Lastly, the number of hyper elements for each case is shown in Table \ref{table:frame_hyper_mesh}.

\begin{table}[h!]
\centering
\begin{tabular}{cc|cc|ccc}
\toprule

         &     & \multicolumn{2}{|c|}{Basis 1 (90 DOFs)}                                                & \multicolumn{2}{c}{Basis 2 (110 DOFs)}                                               \\ 
\midrule
Force [N] & HF [s] & HR [s] & Speedups & HR [s] & Speedups     \\ \midrule
100          & 2761         &   156    & 17.7      &  111      & 18.5                        \\
500           & 6030       &    407   & 14.8      & 501       & 13.2                           \\
1000            & 11329       &    937   & 12.1      & 973     & 12.1                        \\
1500           & 16078       &  1269    & 12.7      & 1335       & 11.9             \\  \bottomrule     
\end{tabular}
\caption{HF and ROM run times for Basis (1) and Basis (2)}
\label{table:frame_online_costs}
\end{table}

\begin{table}[h!]
\parbox{.45\linewidth}{
\centering
\begin{tabular}{c|cc}
\toprule
Offline   cost [s]                                   & Basis (1) & Basis (2) \\
\midrule
Total training   time & 1.38     & 1.76      \\
Basis time                                & 41.63     & 65.1    \\ \bottomrule 
\end{tabular}
\caption{ROM offline costs for Basis (1) and Basis (2)}
\label{table:frame_offline_costs}
}
\hfill
\parbox{.45\linewidth}{
\centering
\begin{tabular}{c|cc}
\toprule
F [N] & Basis (1) & Basis (2) \\
\midrule
100       & 43             & 47             \\
500         & 52             & 57             \\
1000         & 68             & 66             \\
1500        & 68             & 73         \\ \bottomrule   
\end{tabular}
\caption{Hyper mesh for Basis (1) and Basis (2)}
\label{table:frame_hyper_mesh}
}
\end{table}

\section{Conclusions}
\label{sec:conclusions}
We presented a hyper-reduced order model for steady-state analysis of jointed structures. The reduction basis is constructed using the Augmented Jacobian Projection method. This involves solutions to a set of autonomous and forced linear problems, corresponding to different contact states covering the expected range of nonlinearity. After the projection of the dynamical system, we adapt the energy conserving, sampling and weighting method for Hyper Reduction (HR) in the frequency domain and use training vectors that are already available from the construction of the Reduced Order Model (ROM). An error indicator that is based on the HR-ROM solution was presented. The main contributions of this work are:

\begin{itemize}
    \item A HR-ROM that is capable of capturing not only global features such as overall dissipation and stiffness changes with high accuracy, but also local quantities, whose convergence is known to be more challenging, such as the evolution of friction forces along the interface.
    \item An error metric that depends solely on the HR-ROM solution, with reported thresholds that have proved reliable for the different analyses we performed.
    \item Allowing convergence studies with respect to interface mesh size for steady-state analysis. This is possible by virtue of the sparse selection of nonlinear element, using the proposed HR scheme.
\end{itemize}

Regarding the limitations of the method, since the construction of the basis involves the selection of a mode shape of interest, the ROM remains valid across a limited range of frequencies around that corresponding eigenfrequency of the linearized structure. An extension of this range can be sought using linear mode shapes of the off-resonant modes, and further, by considering basis vectors that represent other modes of interest in the reduction basis. 

\paragraph{Declaration of Competing Interest\\}
\addcontentsline{toc}{section}{Declaration of Competing Interest}
The authors declare that they have no known competing financial interests or personal relationships that could have appeared to influence the work reported in this paper.

\paragraph{Acknowledgments\\}
\addcontentsline{toc}{section}{Acknowledgements}
The authors acknowledge the funding of the Swiss National Science Foundation project “\textbf{Meso}-scale modeling of \textbf{Fri}ction in reduced non-linear interface \textbf{Dy}namics: \textbf{MesoFriDy}".

\bibliographystyle{unsrt}
\bibliography{references}

\end{document}

%% file: commands.tex
\renewcommand{\vec}[1]{\boldsymbol{\mathrm{#1}}}     
\newcommand{\vecf}[1]{\boldsymbol{#1}}        
\newcommand{\mat}[1]{\boldsymbol{#1}}           

\renewcommand{\dd}{\mathrm{d}}                    
\newcommand{\pd}{\partial}                      

\renewcommand{\norm}[1]{\| #1 \|}                 
\renewcommand{\abs}[1]{| #1 |}

\newcommand{\mt}[3]{\mat{#1}_\text{#2}^\text{#3}}
\newcommand{\vt}[3]{\vec{#1}_\text{#2}^\text{#3}}
\newcommand{\st}[3]{\text{#1}_\text{#2}^\text{#3}}
